\documentclass[%
reprint,
 amsmath,amssymb,
 aps,
]{revtex4-2}
\usepackage{graphicx}
\usepackage{graphicx}
\usepackage{dcolumn}
\usepackage{bm}

\begin{document}

\preprint{APS/123-QED}

\title{From laminar to chaotic flow via stochastic resonance in viscoelastic channel flow}

\author{Yuke Li}
 \email{yuke.li@weizmann.ac.il}
 \affiliation{Department of Physics of Complex Systems,
Weizmann Institute of Science, Rehovot 7610001, Israel}

\author{Victor Steinberg}
 \email{victor.steinberg@weizmann.ac.il}
 \affiliation{Department of Physics of Complex Systems,
Weizmann Institute of Science, Rehovot 7610001, Israel}

\date{\today}

\begin{abstract}

         Recent research indicates that low-inertia viscoelastic channel flow experiences supercritical non-normal mode elastic instability from laminar to sustained chaotic flow due to finite-size perturbations. The challenge of this study is to elucidate a realization of such a pathway when the intensity of the elastic wave is too low to amplify velocity fluctuations above the instability onset. The study identifies two subregions in the transition flow regime at Weissenberg number $Wi>Wi_c$, the instability onset. In the lower subregion at $Wi_c\leq Wi\leq 300$, we discover periodic spikes in the streamwise velocity time series $u(t)$ that appear in the chaotic power spectrum as low-frequency, high-intensity peaks resembling stochastic resonance (SR). In contrast, the spanwise velocity power spectrum, $E_w$, remains flat with low-intensity, noisy, and broad  elastic wave peaks. The spikes significantly distort the probability density function of $u$, initiating and amplifying random streaks and wall-normal vorticity fluctuations. The SR appearance is similar to dynamical systems where chaotic attractor and limit cycle interact with external white noise. This similarity is confirmed by presenting a phase portrait in two subregions of the transition regime. In the upper subregion at $Wi>400$ the periodic spikes disappear and $E_w$ becomes chaotic with a large intensity elastic wave sufficient to self-organize and synchronize the streaks into cycles and to amplify the wall normal vorticity according to a recently proposed mechanism.
\end{abstract}

\maketitle

\section{Introduction}
The transition from laminar to turbulent flow in Newtonian or to chaotic flow in viscoelastic fluids follows two distinct pathways. One pathway, common to both fluids, is attributed to a linear normal mode instability followed by a sequence of linear instabilities leading to turbulent or chaotic flow \cite{Drazin2004hydrodynamic}. Linear stability analysis has been successfully applied and validated for various flow geometries with curved streamlines  \cite{Drazin2004hydrodynamic,Larson_JFM1990,shaqfeh1996purely,Steinberg_ARFM2021}. It is important to note that this approach is only valid for Hermitian equations (or operators) characterized by normal eigenmodes, which are prone to linear instability. The onset of instability is determined by the critical eigenvalue of the most unstable normal eigenmode, which grows exponentially. Furthermore, nonlinear effects stabilize the critical eigenmode at a sufficiently large amplitude, thereby replacing a laminar flow that is stable at $Wi<Wi_c$. 


In the second scenario, which occurs in both Newtonian and viscoelastic parallel shear flows, turbulent or chaotic flows are induced by finite-sized external perturbations just above the instability threshold and persist under steady conditions. This type of instability, known as non-normal mode instability, occurs in a flow described by non-Hermitian equations characterized by both normal and non-normal modes, as first discussed for Newtonian parallel shear flows \cite{schmid2007nonmodal,trefethen1993hydrodynamic,kerswell2018nonlinear}. The study of the linear instability of Newtonian parallel shear flows uses a linearized version of the Navier-Stokes equation, called the Orr-Sommerfeld equation, a classic example of non-Hermitian equations. Historically, this equation has been used extensively in its Hermitian approximation at Reynolds number $Re\gg 1$, leading to its linear stability at all $Re$ \cite{Drazin2004hydrodynamic}. More recently, its non-Hermitian form has been used to study the instability of Newtonian flows. This approach shows that while stable normal modes decay exponentially, unstable non-normal modes grow algebraically and can be amplified by several orders of magnitude over time \cite{schmid2007nonmodal,trefethen1993hydrodynamic,kerswell2018nonlinear}.

A linear normal mode instability is driven by a deterministic instability mechanism based on the competition between a destabilizing force and a stabilizing factor, such as  dissipation or external fields, especially when the former predominates \cite{cross1993revmodphys,Drazin2004hydrodynamic}. Specifically, in polymer solution flows with curved streamlines, linear instability arises due to \lq\lq hoop stress\rq\rq, an elastic stress generated by polymers stretched along curved streamlines by the velocity gradient. The hoop stress generates a destabilizing bulk force directed toward the center of curvature, which competes with relaxation dissipation of the elastic stress. The onset of linear elastic instability, denoted by the Weissenberg critical number $Wi_c$, is determined by its mismatch \cite{Larson_JFM1990,shaqfeh1996purely,Steinberg_ARFM2021}. Here, $Wi\equiv U\lambda/ d$ is the main control parameter in inertia-less, viscoelastic fluid flow and denotes the ratio of elastic stress, generated by polymer stretching due to its entropic elasticity to stress dissipation due to its relaxation, where $U$ is the mean flow velocity, $d$ is the characteristic vessel size, and $\lambda$ is the longest polymer relaxation time. In the present experiment, we consider $Wi\gg 1$ and the Reynolds number $Re\equiv Ud \rho/\eta \ll 1$, which corresponds to an extremely large elasticity number $El=Wi/Re=\lambda \eta/\rho d^2 >>1$, where $\rho$ and $\eta$ are the solution density and dynamic viscosity, respectively (see Materials and Methods for details). 

However, the instability mechanism becomes ineffective at a zero-curvature limit \cite{Larson_JFM1990,pakdel1996elastic}. Therefore, viscoelastic parallel shear flows have been shown to be linearly stable over all $Wi$ and for $Re\ll 1$ \cite{Leonov_JMM1967,Renardy_JNFM1986}, which is analogous to Newtonian parallel shear flows at all $Re$. For the latter, however, the proven linear stability fails to explain Reynolds’ seminal experiments, where a transition from laminar to turbulent flow was observed in Newtonian pipe flow at finite $Re$ \cite{Drazin2004hydrodynamic,trefethen1993hydrodynamic,morozov2022JNNFM}. Thus, linear stability does not necessarily imply global stability in either Newtonian \cite{Drazin2004hydrodynamic,trefethen1993hydrodynamic} or viscoelastic \cite{jovanovic2010transient,jovanovic2011nonmodal,page2014streak,hariharan2021localized,lieu2013worst,morozov2022JNNFM} parallel shear flows. Similar to Newtonian flows, inertia-less viscoelastic parallel shear flows exhibit non-normal mode instability at $Wi\gg 1$ and $Re\ll 1$ due to the non-Hermitian nature of the linearized elastic stress equation \cite{jovanovic2010transient,jovanovic2011nonmodal,page2014streak,hariharan2021localized,lieu2013worst,morozov2022JNNFM}. These papers focus on theoretical investigations and numerical analyses of the non-normal unstable modes in viscoelastic channel and plane Couette flows only during the linear transient growth at $Wi\gg 1$ and $Re\ll 1$. During the transient growth, the emergence of coherent structures (CSs), such as streamwise streaks and rolls, is observed. However, the linear growth model fails to predict the presence of nonlinearly stabilized sustained CSs. 

The first experiments on viscoelastic parallel shear flows are carried out in pipes \cite{yesilata2002fluiddyn,yesilata2009polymereng,bonn2011large}, then extended to straight square microchannels with strong pre-arranged perturbations at the inlet \cite{pan2013nonlinear,qin2017characterizing,qin2019flow}. These experiments report the finding of significant velocity fluctuations above $Wi_c$, which contradicts the theoretically proven linear stability of such viscoelastic flows \cite{Leonov_JMM1967, Renardy_JNFM1986}. Subsequent experiments performed in our laboratory on viscoelastic straight channel flows at $Wi\gg 1$ and $Re\ll 1$ with different external perturbation intensities \cite{jha2020universal,jha2021elastically,shnapp2022nonmodal, Steinberg_LTP2022,Li2023flowprop,Li2023non-Hermitian}, find a  supercritical instability at $Wi\geq Wi_c$  leading directly to a sustained chaotic flow. This transition is characterized by the dependence of the friction factor, the normalized root mean square (rms) streamwise velocity and pressure fluctuations on $Wi-Wi_c$, with scaling exponents that deviate from the 0.5 typical for normal mode supercritical bifurcation \cite{Drazin2004hydrodynamic,cross1993revmodphys}. Moreover, $Wi_c$ depends on  the amplitude and structure of the external perturbations, in contrast to the normal mode bifurcation \cite{Drazin2004hydrodynamic}. These two key features - the direct transition from laminar to chaotic flow and the dependence of $Wi_c$ on the intensity and spectral characteristics of the external perturbations - validate the non-modal elastic instability \cite{jovanovic2011nonmodal,morozov2022JNNFM,datta2022PRF}. At $Wi>Wi_c$, regardless of the strength and structure of the external perturbations, three chaotic flow regimes emerge: transition, elastic turbulence (ET), and drag reduction (DR), each accompanied by elastic waves just above $Wi_c$. The experimental results show a universal scaling of the flow properties, a dependence of the elastic wave propagation velocity on $Wi-Wi_c$, a power-law decay of the velocity power spectra, and a predominant presence of streaks \cite{jha2020universal,jha2021elastically,shnapp2022nonmodal,Steinberg_LTP2022, Li2023flowprop,Li2023non-Hermitian}. 

The existence of elastic waves in three flow regimes is the characteristic feature of viscoelastic chaotic channel flow. The waves are predicted to appear in the turbulent flow of viscoelastic fluid at $Re\gg 1$ and $Wi\approx 1$, and are expected to be suppressed by large viscous dissipation at $Re\ll 1$ and $Wi\gg 1$ \cite{Lebedev2001PRE}. Contrary to these predictions, the elastic waves were first discovered in the inertia-less viscoelastic flow at $Re\ll 1$ and $Wi\gg 1$ in the ET flow regime. This discovery is characterized by distinct peaks in the spanwise velocity power spectra, with frequency and intensity varying with $Wi$, and in particular by their propagation velocity in the direction of the streamwise velocity, which is characterized by a scaling law as a function of $Wi-Wi_c$, as reported in Ref. \cite{varshney2019NatComm}. Subsequent research has confirmed this phenomenon and established its universality in channel flows with different external noise intensities  \cite{Steinberg_ARFM2021,Steinberg_LTP2022,Li2023non-Hermitian,Li2023flowprop,jha2020universal,jha2021elastically,shnapp2022nonmodal}.  A simple physical explanation for the appearance of elastic waves is based on the analogy of the response of an elastic stress field to transverse perturbations, similar to the response of an elastic string when plucked \cite{Steinberg_ARFM2021}.

Here, the experimental study focuses on the immediate vicinity above the onset of the supercritical elastic instability of inertia-less viscoelastic planar channel flow subjected to finite-size perturbations induced by the unsmoothed inlet and two small cavities in the upper wall at both channel ends. For $Wi\geq Wi_c$, the instability leads to a transition from laminar to sustained chaotic flow. Recent experimental evidence reveals the key role of elastic waves in amplifying wall normal vorticity fluctuations through resonant pumping of elastic wave energy withdrawn from the mean shear flow into the fluctuating wall normal vorticity \cite{Li2023flowprop}. The higher(lower) the elastic wave intensity, the higher(lower) the flow resistance and the rotational vorticity fluctuations \cite{Li2023flowprop}. This mechanism also helps to explain the transition from ET to DR first described in Ref. \cite{varshney2018PRF} and explained in Ref. \cite{kumar2022PRF}. Given the key role of elastic waves in promoting sustained chaotic flow, the main goal of the current experiment is to uncover the pathway to chaotic flow, in particular when the energy of the elastic wave is insufficient to initiate and amplify streaks and wall normal vorticity fluctuations above $Wi_c$. 

\section{Results}

\begin{figure}
\centering
\includegraphics{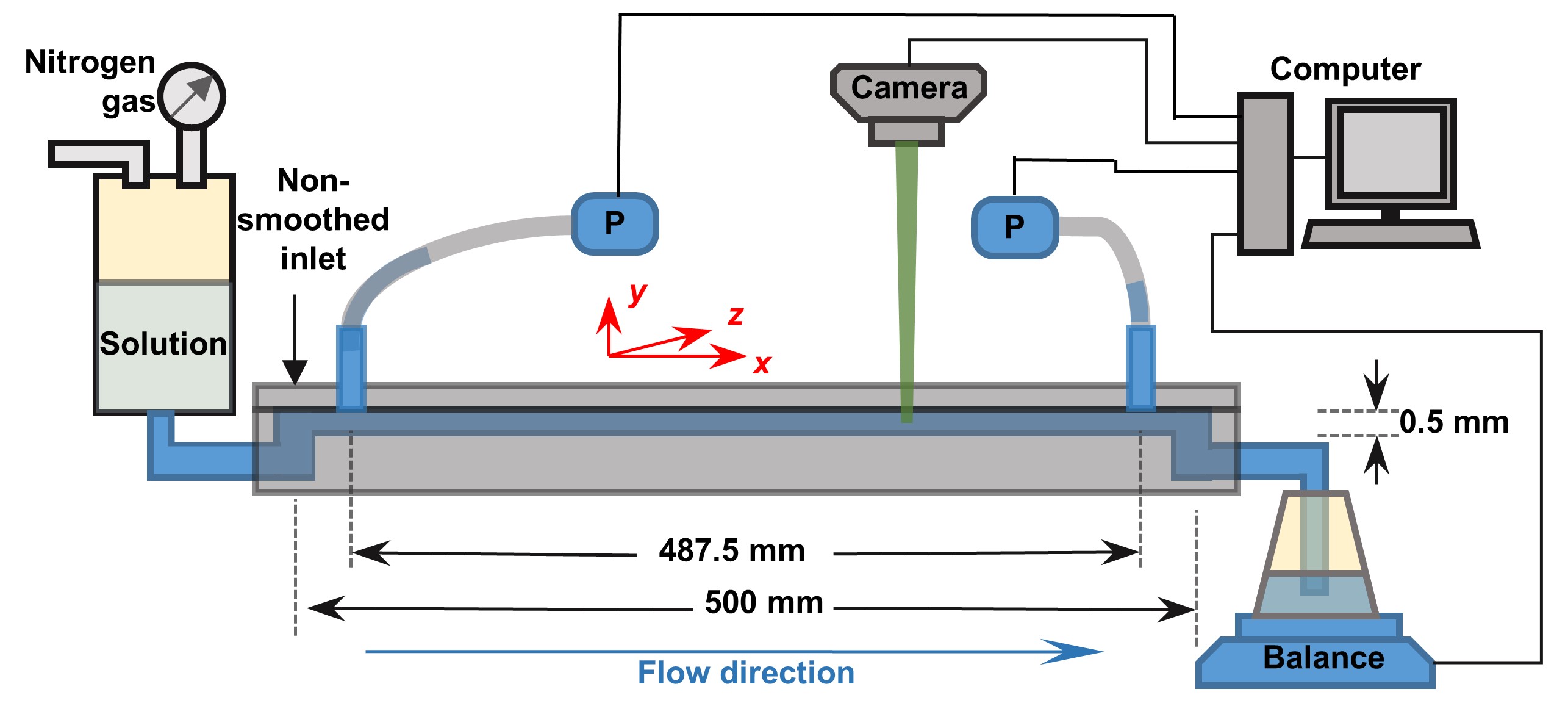}
\caption{Schematic of the experimental setup. Flow in the long straight channel with the non-smoothed inlet and two small cavities in the top plate close to the inlet and outlet, used for absolute pressure measurement, is driven by compressed Nitrogen gas (from left to right in the schematic). Pressure, velocity, and flow discharge are measured by absolute pressure sensors, a high-speed and high spatial resolution camera with laser sheet illumination, and a balance interfaced with a computer, correspondingly.}
\label{fig:1_sch}
\end{figure}

We performed the experiment in a long channel of dimensions 500 ($L$, $x$: streamwise) $\times$ 0.5 ($h$, $y$: wall-normal) $\times$ 3.5 ($W$, $z$: spanwise) mm$^3$,shown in Fig. \ref{fig:1_sch}. The channel has an unsmoothed inlet and two small holes of 0.5 mm diameter in the top plate near the inlet and outlet, used for pressure measurements, which introduce finite-size perturbations, much weaker than strong pre-arranged perturbations used in Refs. \cite{jha2020universal,jha2021elastically} and slightly weaker than in the channel discussed in Ref. \cite{Li2023non-Hermitian}. The latter is determined by the rms of streamwise velocity fluctuations, $u_{rms}$, at the inlet. A dilute aqueous solution with 64\% sucrose concentration and 80 ppm high molecular weight polyacrylamide (PAAm) is the same as used in Refs. \cite{Li2023non-Hermitian, Li2023flowprop}. Its longest relaxation time ($\lambda$)  is 13 s. In the Materials and Methods section, we discuss details of the experimental setup, preparation and characterization of the polymer solution, pressure ($p$) fluctuations, and Particle Image Velocimetry (PIV) measurement techniques.

\subsection{Characterization of the elastic transition at $Wi\geq Wi_c$}
\begin{figure*}
\centering
\includegraphics{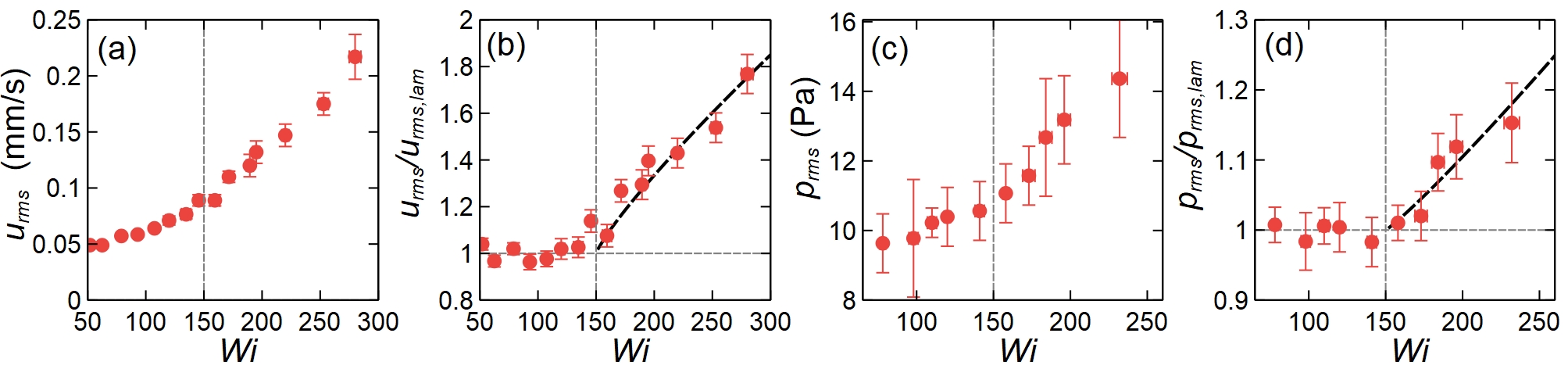}
\caption{Characterization of elastic instability and transition regime as a function of $Wi$ in a limited range above $Wi_c$. (a) rms stream-wise velocity fluctuations, $u_{rms}$, (b) normalized rms stream-wise velocity fluctuations,$u_{rms}/u_{rms,lam}$,  (c) rms pressure fluctuations, $p_{rms}$, (d) normalized rms pressure fluctuations, $p_{rms}/p_{rms,lam}$, versus $Wi$ in linear coordinates. The black dashed line is the power-law fit of $u_{rms}/u_{rms, lam}$ and $p_{rms}/p_{rms, lam}$ versus $(Wi/Wi_c-1)^\epsilon$+1 with $\epsilon=0.85$ and 1.10, respectively, above and in close vicinity to $Wi_c$.}
\label{fig:2_uprms}
\end{figure*}

The onset of the elastic instability at $Wi_c=150\pm10$ and the characterization of the channel flow at $Wi\geq Wi_c$ in three subsequent chaotic flow regimes, namely transition, ET, and DR, are described in Ref. \cite{Li2023flowprop}. These three flow regimes are well characterized by the dependence of the friction factor, normalized pressure, and streamwise velocity fluctuations on $Wi/Wi_c-1$ at the downstream location $l/h=380$, as reported in the same reference \cite{Li2023flowprop}. The corresponding exponents are found to be universal for the studied viscoelastic channel flows with different external perturbations \cite{jha2020universal,shnapp2022nonmodal,Li2023non-Hermitian}. Examining $u_{rms}$ and rms pressure fluctuations, $p_{rms}$, versus $Wi$ below and above $Wi_c$, one finds their continuous growth without any clear indication of the instability onset as shown in Figs. \ref{fig:2_uprms}(a) and \ref{fig:2_uprms}(c). Algebraic fits of the normalized streamwise velocity fluctuations, $u_{rms}/u_{rms, lam}$, and pressure fluctuations, $p_{rms}/p_{rms, lam}$, versus $Wi/Wi_c-1$ up to $Wi=280$ yield exponent values of 0.85 and 1.10, respectively. These values are consistent with those obtained in the previous studies, where the exponents were obtained for data taken in the entire transition regime as shown in Figs. \ref{fig:2_uprms}(b) and \ref{fig:2_uprms}(d) \cite{jha2020universal,shnapp2022nonmodal, Li2023non-Hermitian, Steinberg_LTP2022}. Here $u_{rms, lam}$ and $p_{rms, lam}$ are rms velocity and pressure fluctuations in laminar flow, respectively. 

\subsection{Temporal evolution of stream-wise velocity fluctuations and low-frequency periodic spikes}
\begin{figure*}
\centering
\includegraphics{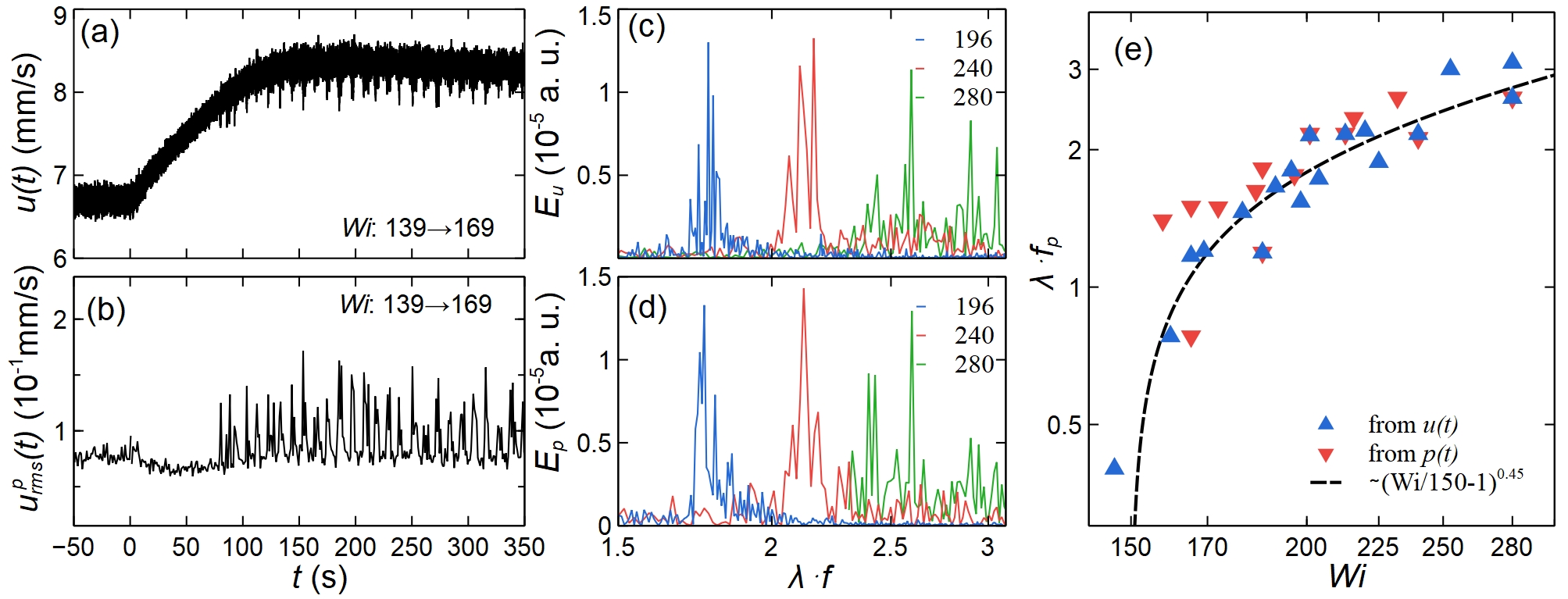}
\caption{Temporal evolution of the streamwise velocity $u(t)$, above $Wi_c=150\pm10$ and low-frequency periodic spikes. (a) Streamwise velocity $u(t)$, and (b) its corresponding partially averaged velocity fluctuations $u^{p}_{rms}(t)$ over each 1s at 100 fps at $l/h=380$ due to a pressure increase at $t=0$ s. The flow varies from $Wi=139$ to $Wi=169$. (c) Streamwise velocity, (d) pressure power spectra at low normalized frequencies, $\lambda f$, in lin-log coordinates at three $Wi$ values. Note the sharp peaks of velocity and pressure fluctuations. (e) $Wi$ dependence of the normalized periodic spikes frequency, $\lambda f_p$, above $Wi_c$  to about the edge of their existence.}
\label{fig:3_spike}
\end{figure*}

\begin{figure*}
\centering
\includegraphics{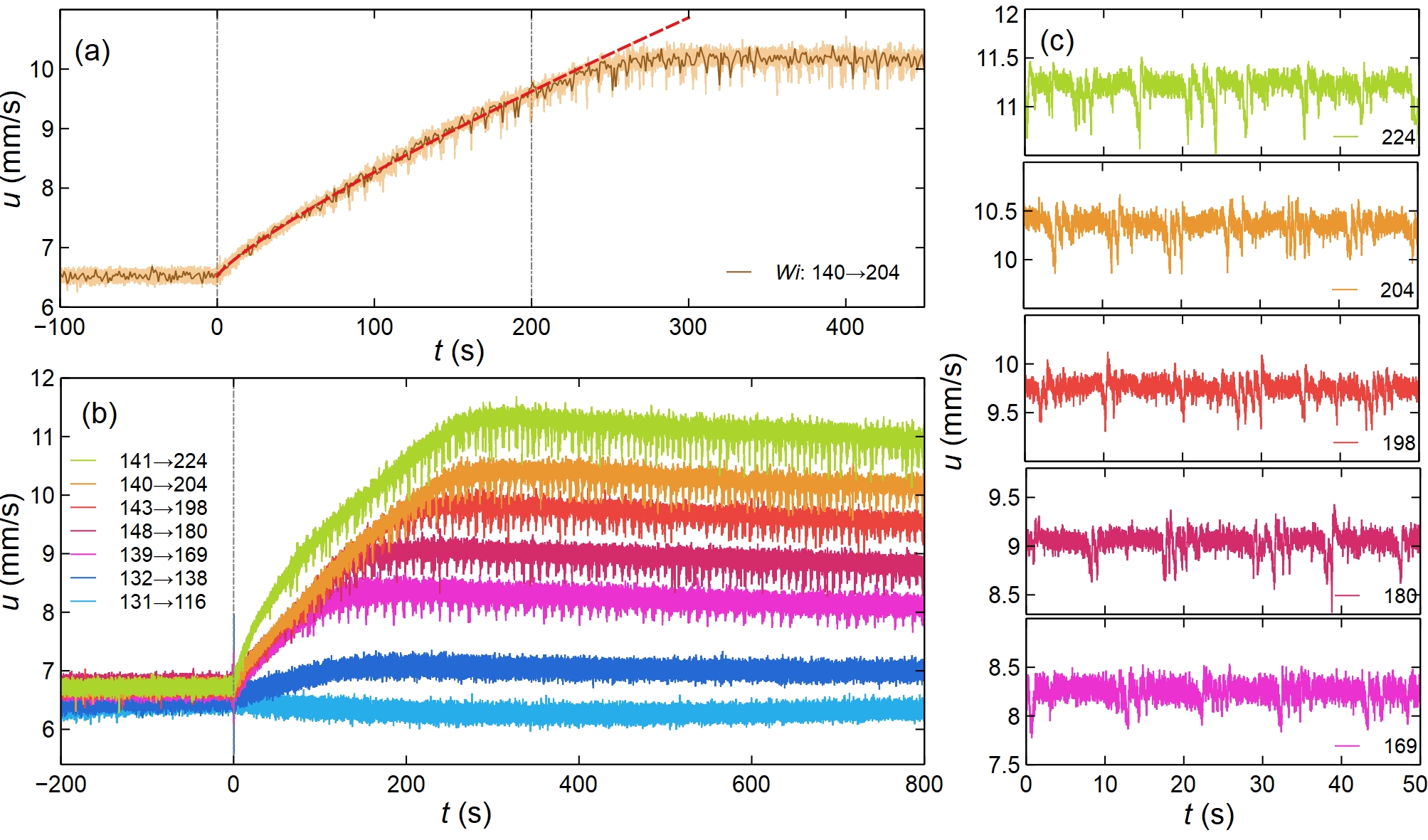}
\caption{Time evolution of velocity fluctuations witnessing a sudden change of inlet pressure. (a) $u(t)$ after pressure increase at the inlet at $t=0$, resulting in $Wi$ growth from 140 to 204. The grey line presents the raw data taken with 100 fps of stream-wise velocity as a function of $t$, $u(t)$, the black presents $u(t)$ averaged over 1 s, and the red dashed line is the fit $A \times t^{\epsilon} + u_0$, with fitting parameters $A=0.0384$, $\epsilon$=0.0829, and $u_0$=6.52 mm/s. (b) Seven cases of $u(t)$ include two scenarios: from laminar to laminar ($Wi$: 131$\rightarrow$116 and 132$\rightarrow$138) and from laminar to transition flow regime at $Wi>Wi_c$ ($Wi$:139$\rightarrow$169, 148$\rightarrow$180, 143$\rightarrow$198, 140$\rightarrow$204, and 140$\rightarrow$224). (c) streamwise velocity at a fixed location as a function of time, $u(t)$, during 50 s at different $Wi=169,180,198,204,224$.  }
\label{fig:4_u(t)}
\end{figure*}

By monitoring the temporal evolution of the velocity at the channel center at $l/h=380$ by PIV, followed by a slight pressure increase at the inlet at $t=0$ s, the streamwise velocity $u(t)$ in laminar flow at $Wi=139$ starts to grow algebraically up to $t\sim100$ s, as shown in Fig. \ref{fig:3_spike}(a). This growth pattern is in stark contrast to the exponential growth in a case of normal mode instability \cite{Drazin2004hydrodynamic}. At $t>100$ s, the growth rate of $u(t)$ decreases sharply, with $u(t)$ reaching a saturation point after $t\approx 150$ s  at $Wi=169>Wi_c$. During the transient growth and subsequent saturation, the streamwise velocity fluctuations, $u'=u-u_{mean}$, show a slight increase, in agreement with the observations in  Figs. \ref{fig:4_u(t)}(a) and \ref{fig:4_u(t)}(b). In particular, at $t\approx 100$ s, large and almost periodic negative spikes appear and persist until the end of time series (Fig. \ref{fig:3_spike}(a)). Here $u$ and $u_{mean}$ are the streamwise velocity and  the mean streamwise velocity averaged over a narrow spanwise extent around the channel centerline, respectively.

To enhance the visibility of these spikes, $u(t)$ is partially averaged over each 1 s interval (100 measurement samples), and the resulting partially smoothed data is plotted as $u^p_{rms}(t)$ in Fig. \ref{fig:3_spike}(b). It can be seen that $u^{p}_{rms}(t)$ shows only a marginal increase at $t>0$ s, with much more pronounced periodic spikes at $t\approx100$ s. In Figs. \ref{fig:3_spike}(c) and \ref{fig:3_spike}(d) of the streamwise velocity power spectra, $E_u$, and pressure power spectra, $E_p$, presented in lin-log coordinates, show narrow spike peaks in the low frequency range for three values of $Wi$ up to 280. However, at $Wi>300$ the energy of these spikes is barely detectable and eventually disappears. Thus, the spikes exist only in the lower subrange of the transition flow regime from $Wi_c$ up to $Wi\approx300$. Finally, the dependence of the normalized spike frequency, $\lambda f_p$, obtained from $E_u$, as a function of $Wi$ is shown in Fig. \ref{fig:3_spike}(e) along with a fitted relation $\lambda f_p\sim (Wi/Wi_c-1)^{0.45}$.

\subsection{Statistical properties of velocity fluctuations}

\begin{figure*}
\centering
\includegraphics{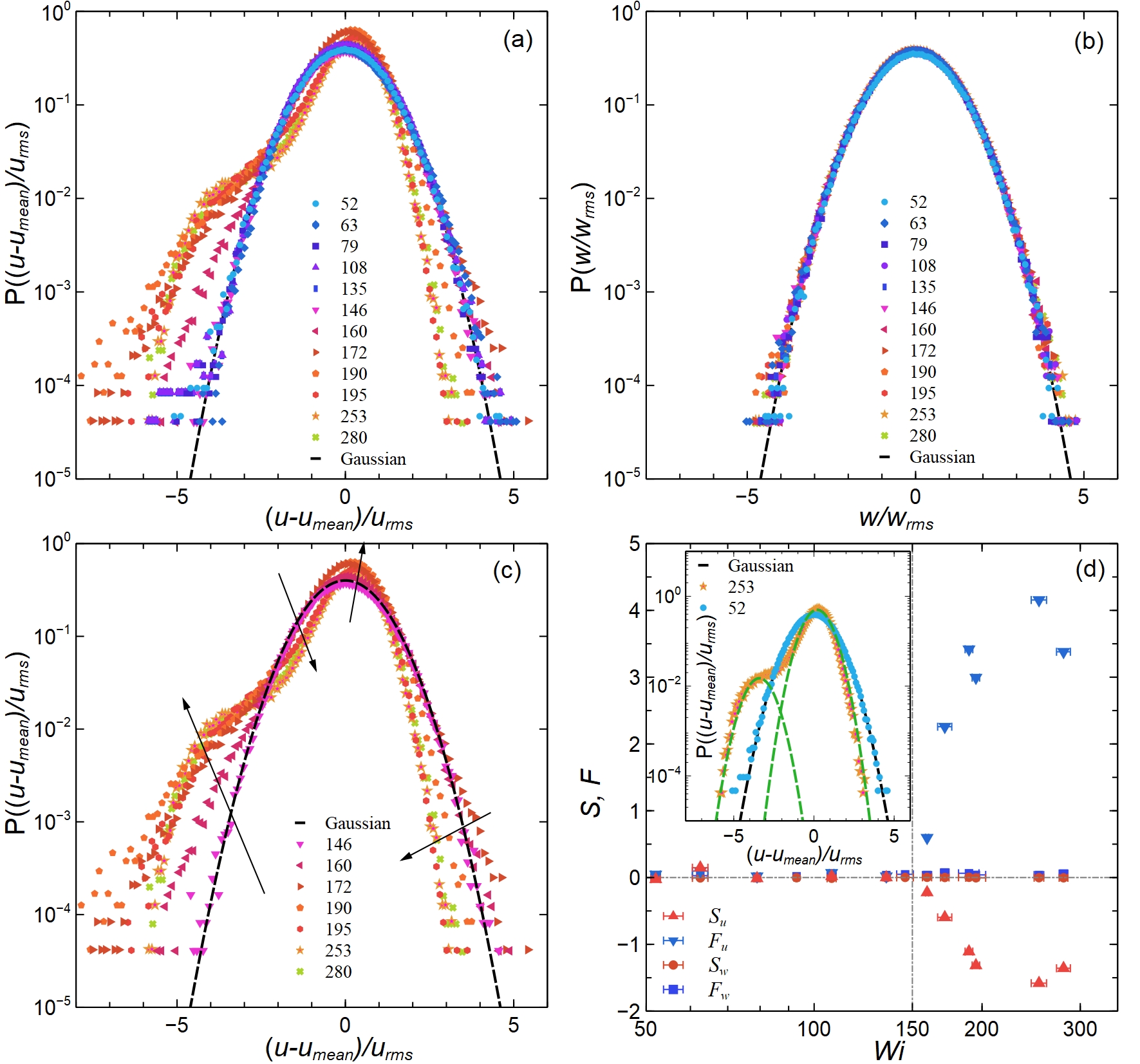}
\caption{PDFs of normalized stream-wise $P((u-u_{mean})/u_{rms})$ and span-wise $P(w/w_{rms})$ velocity fluctuations at the channel center and various $Wi$ values below and above $Wi_c$. (a) $P((u-u_{mean})/u_{rms})$ and (b) $P(w/w_{rms})$ in the whole lower sub-range of $Wi$, (c) $P((u-u_{mean})/u_{rms})$ only above $Wi_c$, and (d) skewness, $S$, and flatness (kurtosis), $F$, of $P((u-u_{mean})/u_{rms})$ and $P(w/w_{rms})$, respectively. Inset: $P((u-u_{mean})/u_{rms})$ for $Wi=52<Wi_c$ and $Wi=253>Wi_c$. The dashed black line is a Gaussian fit to the data at $Wi<Wi_c$, and dashed green lines are Gaussian partial fits to the data at $Wi>Wi_c$ with extensions to negative and positive values of $(u-u_{mean})/u_{rms}$, respectively. }
\label{fig:5_uPDF}
\end{figure*}

\begin{figure*}
\centering
\includegraphics{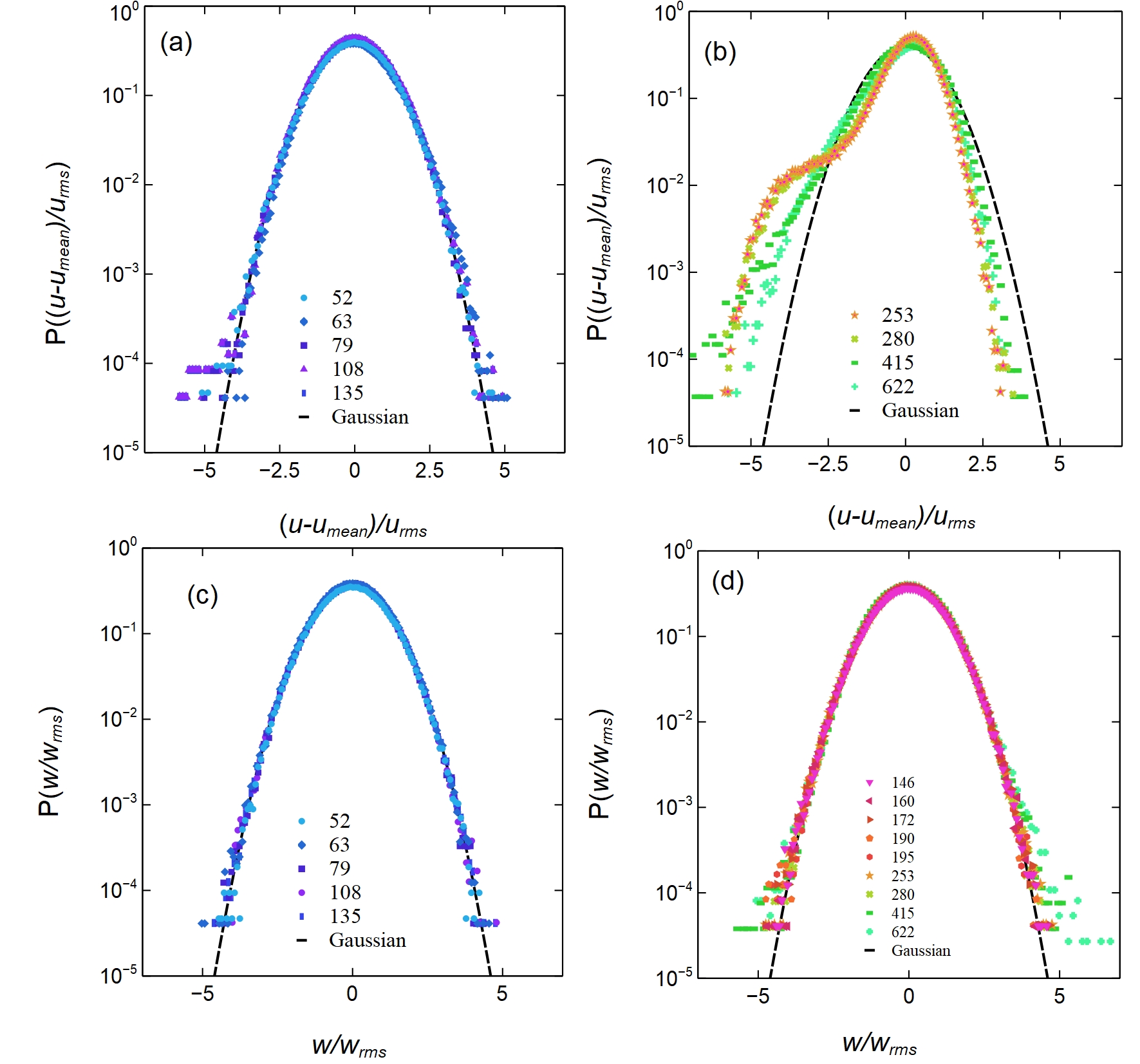}
\caption{PDFs of spanwise and streamwise velocities at channel center. PDFs of $(u-u_{mean}/u_{rms})$ at (a) $Wi<Wi_c$, and (b) $Wi>Wi_c$. PDFs of $w/w_{rms}$ at (c)  $Wi<Wi_{c}$,  and (d) $Wi>Wi_{c}$.   }
\label{fig:6_wPDF}
\end{figure*}

Figure \ref{fig:5_uPDF}(a) shows the probability density functions (PDFs) of normalized streamwise velocity fluctuations, $P((u-u_{mean})/u_{rms})$, over a range of $Wi$ from 52 to 280 below and above $Wi_c$. In particular, the first deviations from Gaussian PDFs appear for negative values of $(u-u_{mean})/u_{rms}$ at $Wi=160$, just above $Wi_c$, with these deviations becoming more pronounced up to $Wi=280$. This is highlighted in Fig. \ref{fig:5_uPDF}(c) by showing the PDFs for $Wi\geq Wi_c$, with black arrows indicating the primary changes from the Gaussian PDF as $Wi$ increases. As can be seen, the PDFs of the positive fluctuations decrease, while the PDFs of the negative fluctuations increase significantly.  This is due to the spikes at lower $u(t)$ values, which are smaller than $u_{mean}$ at $Wi>Wi_c$ (Figs. \ref{fig:3_spike}(a) and \ref{fig:4_u(t)}). The inset in Fig. \ref{fig:5_uPDF}(d) compares PDFs at $Wi=52<Wi_c$ and $Wi=253>Wi_c$, where the latter is fitted by two Gaussian PDFs to emphasize the distortion in the PDF. The main plot in Fig. \ref{fig:5_uPDF}(d) illustrates the abrupt changes in skewness ($S$) and flatness (kurtosis) ($F$) corresponding to the third and fourth moments of $P((u-u_{mean})/u_{rms})$ at $Wi>Wi_c$. Remarkably, their deviations from zero precisely determine the instability onset at $Wi_c=150$. In contrast, the PDFs of the normalized spanwise velocity fluctuations, $P(w/w_{rms})$, show only small deviations from the Gaussian distribution within the error bars in Fig. \ref{fig:5_uPDF}(b) for the same $Wi$ range (Figs. \ref{fig:6_wPDF}(c) and \ref{fig:6_wPDF}(d)). Moreover, their $S$ and $F$ remain zero both below and above $Wi_c$ (Fig. \ref{fig:5_uPDF}(d)).

\subsection{Streamwise and spanwise velocity power spectra and elastic waves}

\begin{figure*}
\centering
\includegraphics{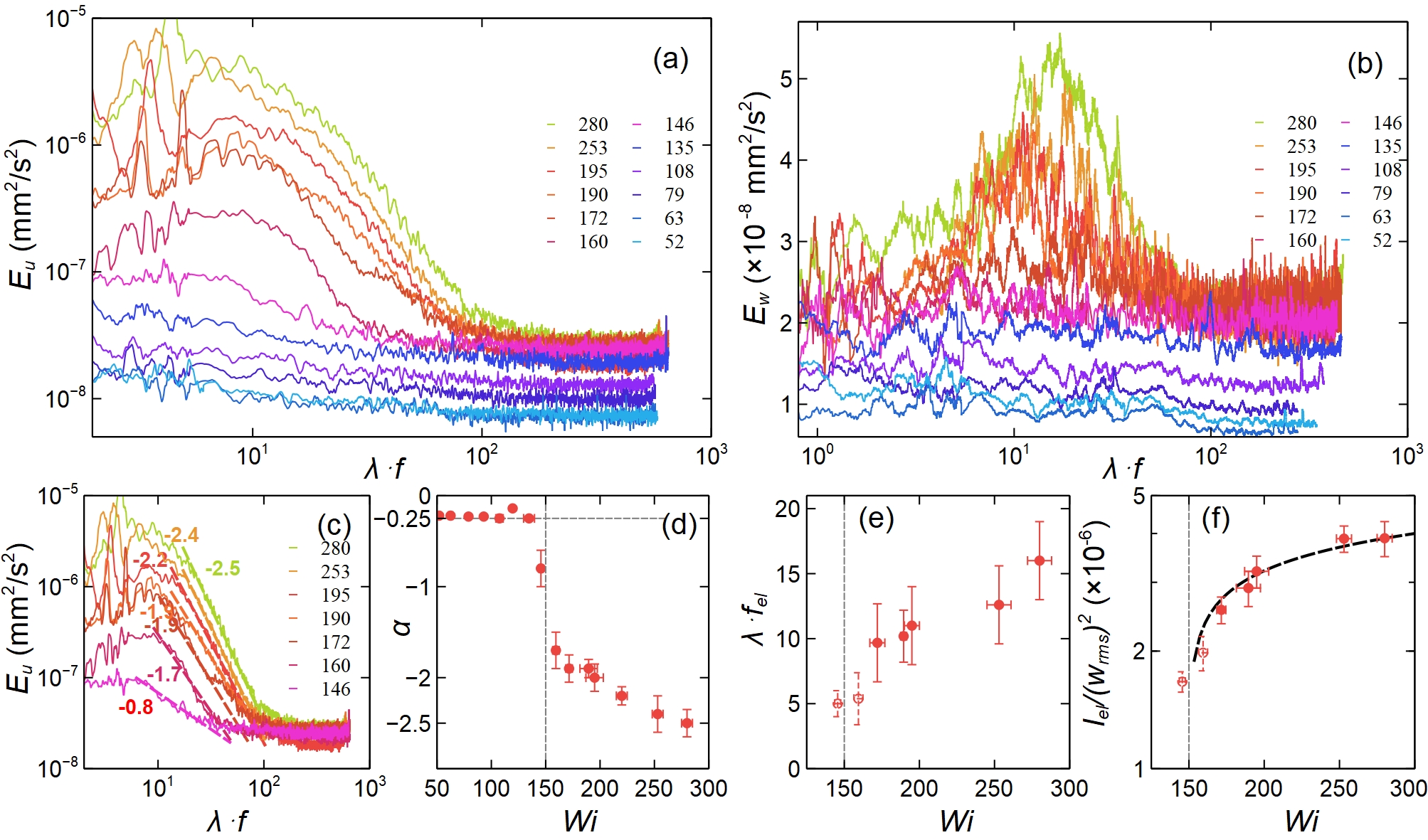}
\caption{Energy spectra and properties of elastic waves. (a) $E_u$ and (b) $E_{w}$ vs. $\lambda f$ at different $Wi$ from 52 to 280 in log-log and lin-log scales, respectively. (c) $E_{u}$ vs. $\lambda f$ at $Wi>Wi_{c}$ with power-law decays fitted by dashed lines with corresponding values of the decay exponents $\alpha$ and colors. (d) Dependence of $\alpha$ on $Wi$. (e) Dependence of the normalized elastic wave frequency, $\lambda f_{el}$, on $Wi>Wi_c$. (f) Dependence of the normalized elastic wave energy, $I_{el}/w_{rms}^2$, on $Wi$ at $Wi>Wi_c$.  The dashed black line is the fit, which gives a scaling relation: $I_{el}/w_{rms}^2 \sim (Wi/Wi_c-1)^{0.2}$. In both (e) and (f) plots the last two points near the onset reach the experimental resolution limit. }
\label{fig:7_uspec}
\end{figure*}

At $Wi<Wi_c$, the streamwise velocity power spectrum, $E_u$, shows a slight increase from nearly flat to slight growth toward lower normalized frequencies, $\lambda f$, reminiscent of a white noise spectrum (Fig. \ref{fig:7_uspec}(a)). However, above $Wi_c$ up to $Wi=280$, $E_u$ shows a significant increase, up to three orders of magnitude at $\lambda f\leq 10$ (Figs. \ref{fig:7_uspec}(a) and \ref{fig:7_uspec}(c)). This increase is accompanied by a sharp power-law decay at $\lambda f\geq 10$, with the exponent $\alpha$ ranging from -0.8 to -2.5 (see Fig. \ref{fig:7_uspec}(d)). Here $\alpha$ is denoted as the exponent of the power-law fit of $E_u$ to $\lambda\cdot{f}$ in the decay regime. 

\begin{figure}
\centering
\includegraphics{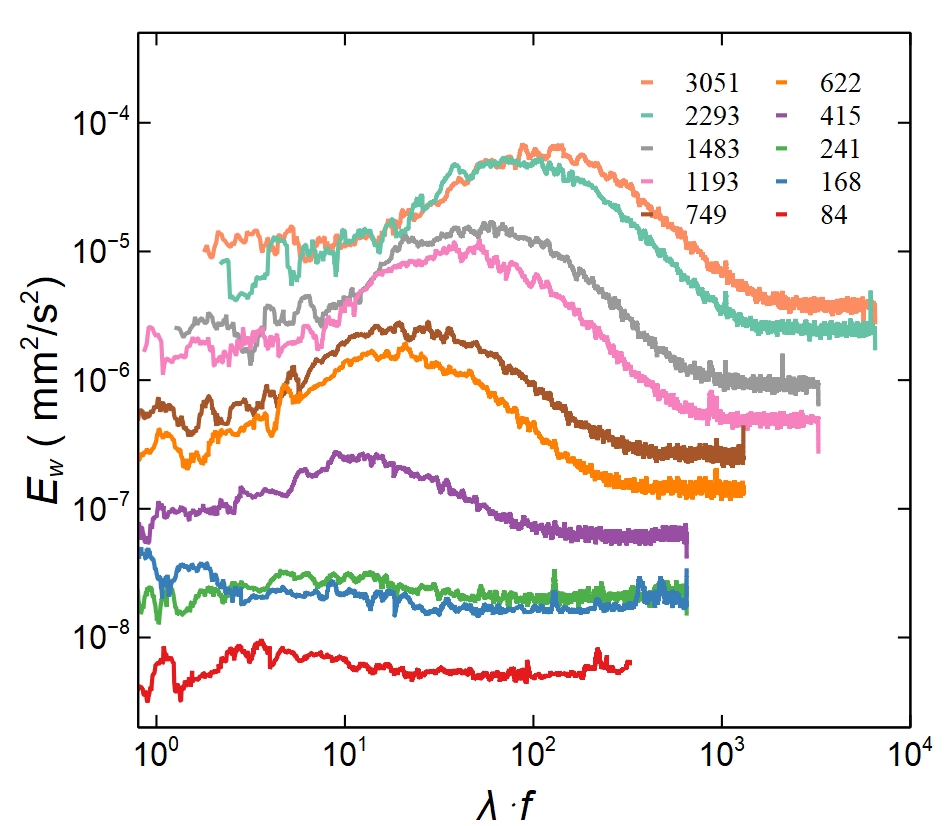}
\caption{Energy spectra of spanwise velocity, $E_w$, versus normalized frequency, $\lambda\cdot f$ in log-log scales
 at $Wi<Wi_c$ in laminar flow and up to $Wi=3051$ in transition, ET and DR flow regimes. It is clearly notable that in the lower sub-range of the transition flow regime elastic wave peaks are not noticeable and an order of magnitude smaller than in the upper sub-range at $Wi=415$ in log-log coordinates.  }
\label{fig:8_wspec}
\end{figure}

In the lower subrange of $Wi>Wi_c$ up to $Wi=280$, the spanwise velocity power spectra, $E_w$, plotted on a linear-logarithmic scale in Fig. \ref{fig:7_uspec}(b) and on a log-log scale in  Fig. \ref{fig:8_wspec}, remain flat and similar to those observed at $Wi<Wi_c$, which characterizes white noise. However, at $Wi>Wi_c$, pronounced broad and noisy peaks associated with the elastic waves appear at the top of the flat power spectra (Fig. \ref{fig:7_uspec}(b)) and are visible in Fig. \ref{fig:8_wspec}.

\begin{figure*}
\centering
\includegraphics{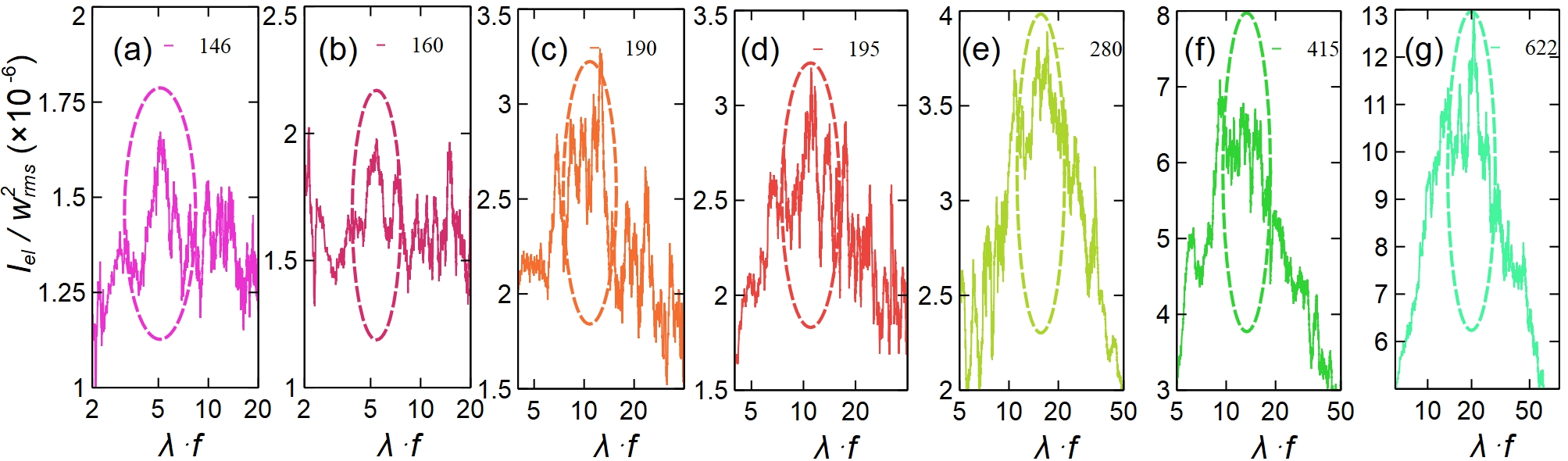}
\caption{Individual looks of elastic waves. $I_{el}/w_{rms}^2$ at $Wi$=146, 160, 172, 190, 195, 253, and 280. Each dashed circle with the corresponding color indicates the appearance of one significant peak.}
\label{fig:9_wave}
\end{figure*}

Since the identification of the peak maximum location and the estimation of its height become unambiguous near $Wi_c$ due to a drastic increase in peak widths and fluctuations as $Wi$ approaches $Wi_c$, the resolution of both $\lambda f_{el}$ and the normalized intensity of the elastic waves, $I_{el}/w_{rms}^2$, is limited to about $\sim 5$ and $\approx 2\times 10^{-6}$, respectively (Figs. \ref{fig:7_uspec}(e) and \ref{fig:7_uspec}(e)(f)). Isolated single peaks in $I_{el}/w_{rms}^2$ versus $\lambda f$ coordinates for $Wi>Wi_c$ provide better resolution (Fig. \ref{fig:9_wave}). In addition,  Fig. \ref{fig:7_uspec}(f) includes a power-law fit of the normalized elastic wave energy dependence on $Wi$ as $I_{el}/w_{rms}^2\sim (Wi/Wi_c-1)^{\beta}$ with $\beta=0.20$.
Remarkably, in the upper subrange of the transition regime at $400\leq Wi< \approx 1000$, and further in the ET and DR regimes, the elastic wave peaks in $E_w$ grow by up to four orders of magnitude, as shown in Fig. \ref{fig:8_wspec}. The location of the peak maximum shifts more than tenfold to higher values of $\lambda f$, and a power-law decay occurs with the scaling exponent increasing with $Wi$. This observation reveals a clear distinction between two subranges in the transition flow regime.

\subsection{Observation of random streaks}

One of the striking findings in inertia-less viscoelastic channel flows, especially when subjected to external perturbations of different amplitudes at $Wi>Wi_c$, is the phenomenon of cyclic self-organized streaks synchronized by elastic waves \cite{Li2023non-Hermitian}. To verify the streak synchronization by the elastic waves we explore the approach based on the velocity difference across the counter-propagating streak interface as a function of normalized time, $t^{*}=tf_{el}$, developed in our laboratory and used in several experiments \cite{jha2020universal,jha2021elastically,shnapp2022nonmodal, Li2023non-Hermitian, Steinberg_LTP2022}. In all these experiments, at $Wi>Wi_c$, we discovered cycling with $t^{*}\simeq 1$ in all three flow regimes.

\begin{figure*}
\centering
\includegraphics{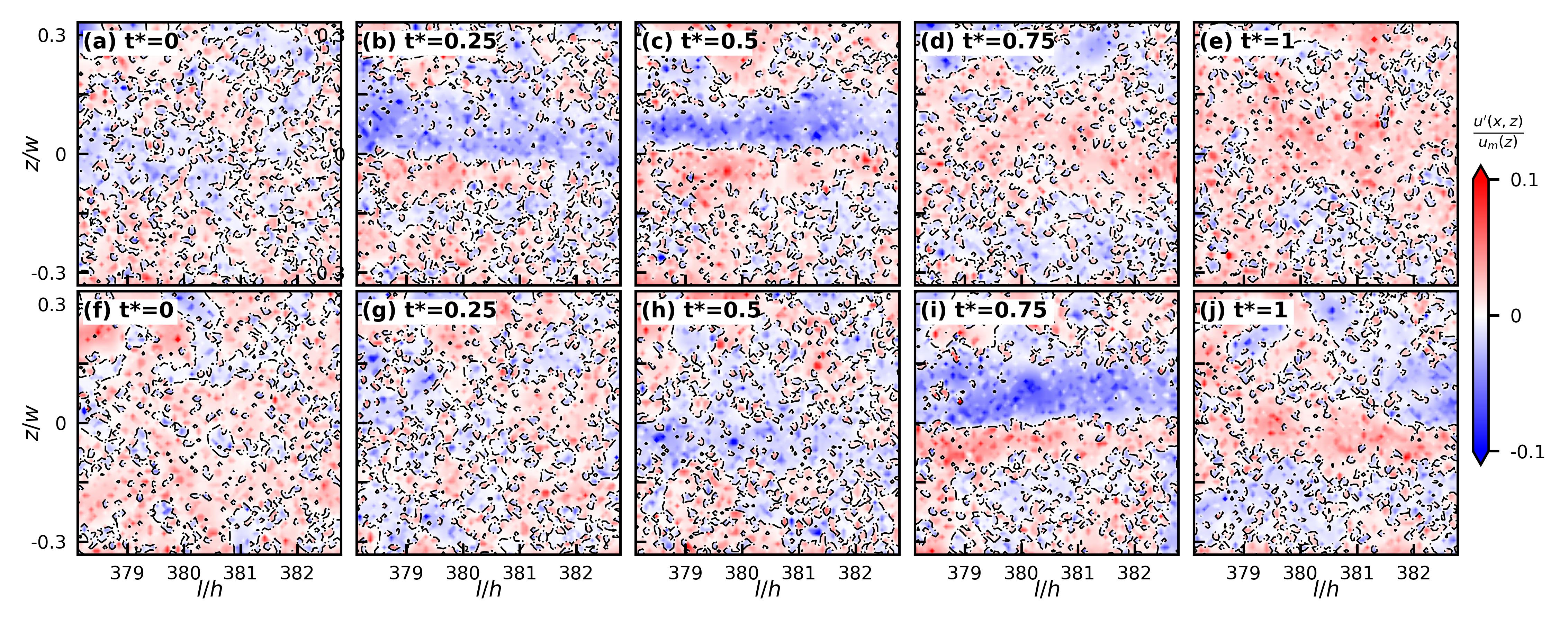}
\caption{Instantaneous images of streamwise velocity fluctuations obtained from the full streamwise velocity $u$ by first subtracting the mean streamwise velocity profile, $u_{mean}(z)$, and then divided by it. Thus, the normalized streamwise velocity fluctuations, $u'(x,z)/u_{mean}(z)$, at $Wi=193$ (upper 5 images) and $Wi$=207 (lower 5 images) for 5 normalized times  $t*=tf_{el}$ are presented for each $Wi$. PIV measurements present images in  $\Delta z/W=0.67$ and $\Delta l/h=4$ spatial window. }
\label{fig:10_streak1}
\end{figure*}

\begin{figure}
\centering
\includegraphics{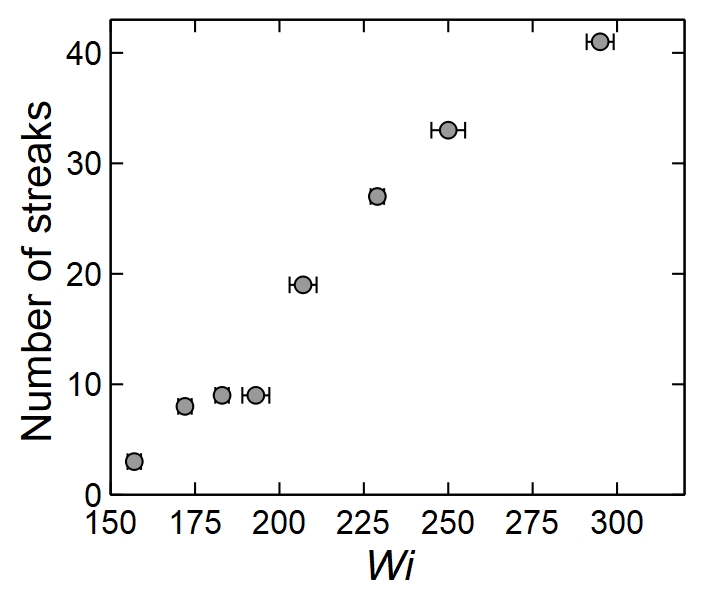}
\caption{Number of streaks observed within 70 seconds as a function of $Wi$. The appearance of streaks is identified by a contour plot of streamwise velocity fluctuations with high resolution via PIV. The color bar of streaks is presented for $Wi>Wi_c$ in the lower sub-range of transition flow regime at $Wi=193$ and 207. }
\label{fig:11_streak2}
\end{figure}

On the contrary, in the lower subrange at $Wi>Wi_c$  up to $Wi=280$, we reveal the emergence of random streaks shown in Fig. \ref{fig:10_streak1} in two rows of 5 consecutive images each, illustrating the streaks as a function of $t^{*}$ at $Wi=193$ and 207. At $Wi\leq 280$ only random streaks appear, although the number of streaks increases significantly with $Wi$ (Fig. \ref{fig:11_streak2}). These streaks lack the organized, cyclic nature observed at higher $Wi$ values. At $Wi\geq 400$, the cycling of self-organized streaks synchronized by the elastic waves, whose energy increases by orders of magnitude, is in agreement with our previous experiments \cite{jha2020universal,jha2021elastically,shnapp2022nonmodal,Li2023non-Hermitian, Steinberg_LTP2022,Li2023flowprop}.    

\section{Discussion}

\begin{figure*}
\centering
\includegraphics{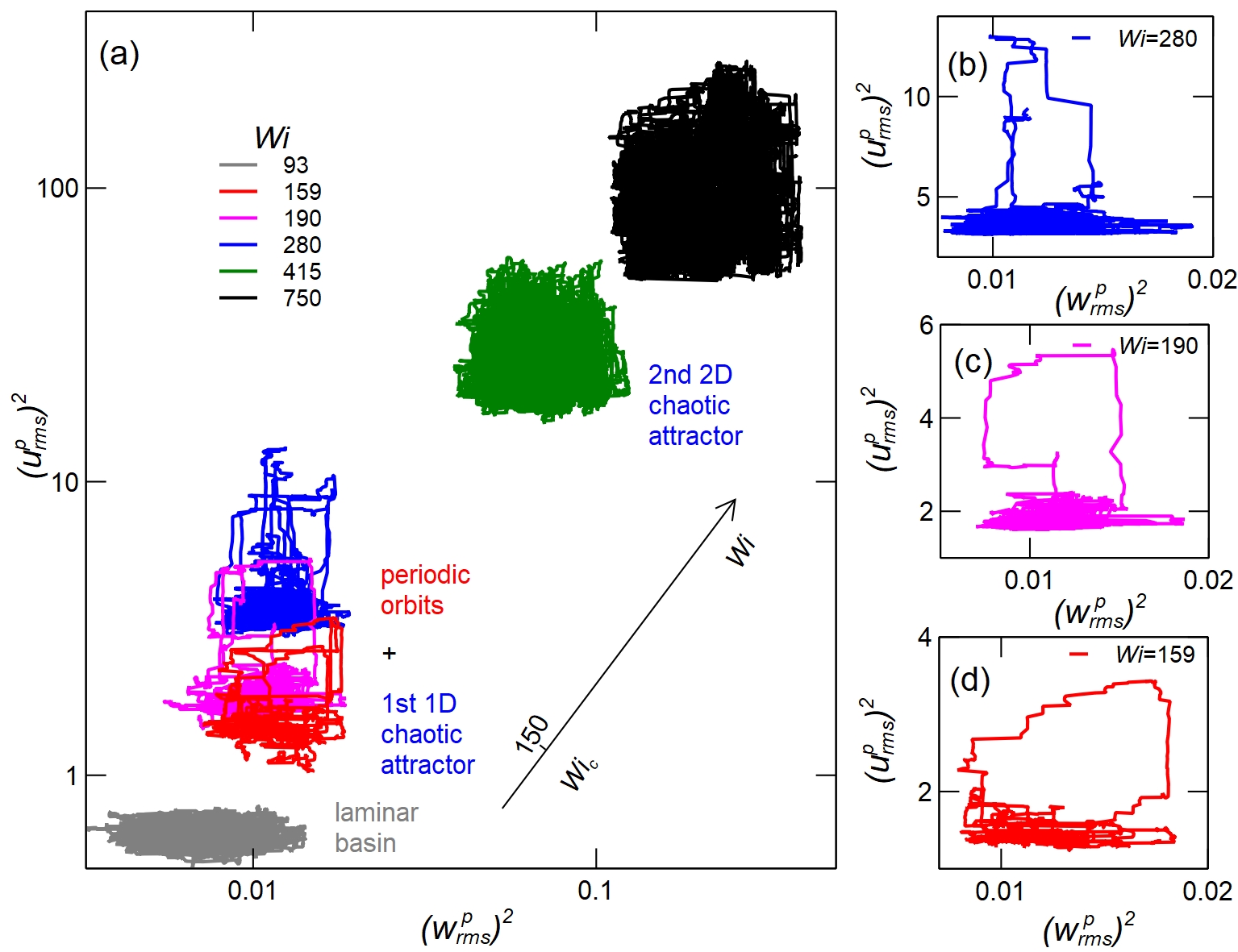}
\caption{Phase portraits of the instantaneous dynamics with increasing $Wi$. (a) $(u^{p}_{rms})^2$ versus $(w^{p}_{rms})^2$ at six $Wi$ values representing three states: (i) laminar basin ($Wi$=93); (ii) chaotic state of $(u^{p}_{rms})^2$ along with periodic orbits and white noise of $(w^{p}_{rms})^2$ at $Wi$=159, 190, and 280; and (iii) the pure chaotic state of both $(u^{p}_{rms})^2$ and $(w^{p}_{rms})^2$ at $Wi$=415 and 750. $(u^{p}_{rms})^2$ and $(w^{p}_{rms})^2$ are computed from squared velocity fluctuations averaged over 1 second (100 points). Individual views of state (ii) are shown separately at (b) $Wi$=280, (c) $Wi$=190, and (d) $Wi$=159.  }
\label{fig:12_phase}
\end{figure*}

This experimental study attempts to elucidate the mechanism behind the supercritical, non-normal mode elastic instability in viscoelastic channel flow with negligible inertia initiated by external finite-size perturbations due to an unsmoothed inlet and two small holes, in the narrow range from $Wi\sim 60$ to $Wi\sim 300$ close to $Wi_c=150$.

The first finding of this study is the existence of two subranges in the transition flow regime above $Wi_c$ with different flow properties, as detailed in the Results section. However, the most surprising discovery is the observation of low-frequency periodic spikes in $u(t)$ and $p(t)$ detected exclusively in the lower subrange at $Wi_c \leq Wi\leq 300$ (Figs. \ref{fig:3_spike} and \ref{fig:4_u(t)}). The periodic spikes appear as sharp low-frequency peaks in the streamwise velocity, $E_u$, and pressure, $E_p$, power spectra presented in lin-log coordinates in Figs. \ref{fig:3_spike}(c) and \ref{fig:3_spike}(d), and only $E_u$ in log-log coordinates in Figs. \ref{fig:7_uspec}(a) and \ref{fig:7_uspec}(c). In the latter, the low-frequency $\lambda f<1$ sharp peaks appear on top of the streamwise velocity fluctuation power spectrum characterized by a power-law decay at $\lambda f>1$, indicating chaotic flow in the streamwise direction. In contrast, the power spectrum of the spanwise velocity fluctuations, $E_w$, slowly increases at $Wi>Wi_c$ in the lower subrange up to $Wi\approx 300$, where $E_w$ still remains flat, similar to that observed at $Wi<Wi_c$, indicating a white noise spectrum in the spanwise direction (Figs. \ref{fig:7_uspec}(b) and \ref{fig:8_wspec}). Thus, in the lower subrange, the chaotic streamwise velocity fluctuations coexist and interact with the spanwise white noise velocity fluctuations in the presence of low-intensity elastic waves. Furthermore, the periodic spikes presented and characterized in Results  are reminiscent of stochastic resonance (SR), a phenomenon considered in autonomous dynamical chaotic systems interacting with external white noise in the presence of weak periodic modulation. In dynamical chaotic systems, the strange attractors found in practice are structurally unstable. This means that such a quasi-attractor exists as several regular chaotic attractors coexisting in the phase space of the system, which merge into one chaotic set at the critical value of the control parameter, the so-called  attractor crisis \cite{anishch1999phys_uspekhi}. Above the crisis, the merged attractor shows intermittent switching of the ``chaos-chaos" type, where the "deterministic stochastic attractor" is observed  \cite{anishch1993jstatphys,anishch1999phys_uspekhi}. The dynamical intermittent switching is a result of additive noise depending on its intensity, which can be synchronized by weak periodic modulation resulting in SR \cite{anishch1993jstatphys,anishch1999phys_uspekhi}. However, the deterministic SR is fundamentally different from the classical SR realized in a bistable system driven simultaneously by noise and a weak periodic signal  \cite{benzi1981JPhys}. To further substantiate the connection of the observations with  deterministic SR, we characterize the flow dynamics by plotting the phase portrait in coordinates of the streamwise, $(u^p_{rms})^2$, versus spanwise, $(w^p_{rms})^2$, velocity energies in the two subranges of the transition flow regime (Fig. \ref{fig:12_phase}(a)). In the lower subrange above $Wi_c$ we find a one-dimensional streamwise velocity chaotic attractor interacting with spanwise velocity white noise and weak elastic waves, resulting in a SR periodic orbit at $Wi=159$, 190, and 280 (Figs. \ref{fig:12_phase}(b)-\ref{fig:12_phase}(d)). We note that in the lower subrange, the energy of the external white noise perturbations $(w^p_{rms})^2$ at $Wi\leq Wi_c$ increases slightly up to about 50\%, while the chaotic streamwise velocity energy, $(u^p_{rms})^2$, increases about tenfold. In contrast, in the upper subrange we detect a two-dimensional chaotic attractor of both $(u^p_{rms})^2$ and $(w^p_{rms})^2$ that grows by more than two orders of magnitude compared to external perturbations in a laminar flow (Fig. \ref{fig:12_phase}(a)). 

\begin{figure}
\centering
\includegraphics[width=0.5\textwidth]{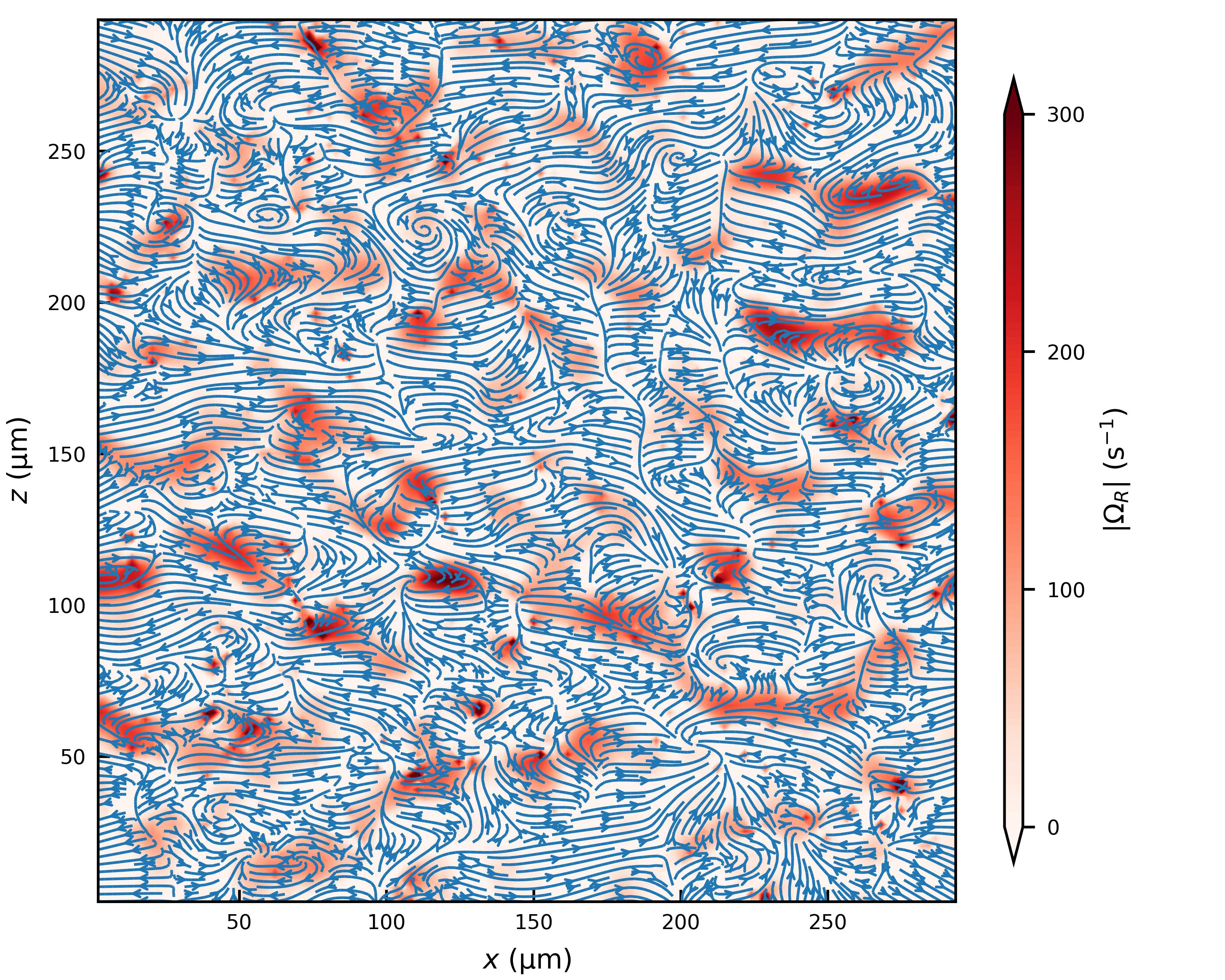}
\caption{An instantaneous snapshot of the 2D image of streamlines of velocity fluctuations field $(u', w')$ in $x$–$z$ plane at the channel center overlapped with the field of the absolute value of the vortical part of the wall-normal vortex fluctuations, $\lvert\Omega_{R}\rvert$, at $Wi$=210. The fluid is seeded with red-fluorescent 0.2 $\mu$m traces of $\sim$0.15\% w/w concentration (Thermo Scientific). The resolution is 512$\times$512 pxl$^2$ at 4000 fps. The PIV experimental setup is the same as that in Fig. 2A in Ref. \cite{Li2023flowprop}.  The color bar of the vortical part of vorticity fluctuations is shown on the right. }
\label{fig:13_vorticity}
\end{figure}

Moreover, in the lower subrange above $Wi_c$, the presence of periodic spikes leads to significant deviations from Gaussian distributions in the PDFs of $(u-u_{mean})/u_{rms}$ at the channel center (Figs. \ref{fig:5_uPDF}(a), \ref{fig:5_uPDF}(c), \ref{fig:5_uPDF}(d), and \ref{fig:6_wPDF}). Other notable features in the lower sub-range include the abrupt change from zero to high absolute values of the PDFs' skewness, $S$, and flatness (kurtosis), $F$, from $Wi_c=150$ to $Wi<300$; the appearance of random streaks (Fig. \ref{fig:11_streak2}); and the observation of wall-normal vortex fluctuations already in the lower subrange (Fig. \ref{fig:13_vorticity}). However, the synchronization at $f_{el}$ of the self-organized cycling streaks and the enhancement of the wall-normal vortex fluctuations occur only in the upper subrange, at $Wi\geq 400$, where the elastic wave energy, $I_{el}$, increases by orders of magnitude compared to the lower subrange (Figs. \ref{fig:7_uspec}(b) and \ref{fig:8_wspec}).

It is also unexpected that the wall-normal vortex fluctuations are likely generated and supported by strong periodic spikes in the lower subrange, while the elastic waves take over their amplification only in the upper subrange at $Wi\geq 400$ \cite{Li2023flowprop}. Remarkably, at $Wi> 300$ the low-frequency periodic spikes disappear, the strongly distorted negative tail of the PDF of $(u-u_{mean})/u_{rms}$ becomes exponential, and the flow becomes chaotic in both the $u$ and $w$ velocity fields. The flow dynamics at $Wi\geq 400$ are characterized by a two-dimensional chaotic attractor, and the scaling of the flow properties is found to be the same as in the transition flow regime of viscoelastic channel flows with different perturbation intensities, where the measurements are made with lower resolution and further away from $Wi_c$ \cite{Li2023non-Hermitian, Li2023flowprop}.

In conclusion, the presented experimental results reveal novel and unexpected features of the non-modal instability evolution above and near $Wi_c$ and provide strong evidence for the stochastic nature of the instability mechanism. Above $Wi_c$, the low-frequency periodic spikes observed in the streamwise velocity time series, characterized by the sharp peaks in $E_u$ in the lower subrange at $Wi_c\leq Wi\leq300$ of the transitional flow regime, are reminiscent of SR. Based on the experimental phase portrait of the instantaneous dynamics, it is proposed that SR arises due to the interaction of streamwise velocity chaotic flow with  spanwise velocity white noise  in the presence of low-intensity elastic waves. This scenario is similar to the deterministic SR found in autonomous chaotic nonlinear dynamical systems, where deterministic SR is generated due to the interaction of a chaotic attractor with external white noise in the presence of a weak periodic signal at the fixed value of the control parameter \cite{anishch1993jstatphys,anishch1999phys_uspekhi}. SR greatly increases the probability of slow streamwise velocity fluctuations and promotes the generation of wall-normal vorticity fluctuations. Moreover, since the elastic waves have low intensity at $Wi_c\leq Wi\leq 300$, which is insufficient to initiate and synchronize the streaks and amplify the wall-normal vorticity fluctuations, SR takes over the role of the elastic waves in initiating the random streaks and amplifying the wall-normal vorticity fluctuations.

The discovery of SR in the limited range of $Wi_c\leq Wi\leq 300$ is crucial for understanding the pathway above supercritical non-modal elastic instability to sustained chaotic flow. This phenomenon may have broader implications for various flows, including those in stress fields, such as magneto-hydrodynamic \cite{schek2022jplasmaphys}, viscoelastic electrolyte \cite{koch2019PRL,das2022jnnfm}, viscous solution of rods \cite{dario2017PRE} and suspension of flexible fibers \cite{legal2014PRL}, active nematics \cite{julia2018natcomm} and bacterial and active turbulence \cite{joanny2022arcmp} in parallel shear flows.  

Finally, we would like to point out similarities and differences with the widely studied theoretically, numerically, and experimentally  Newtonian pipe flow at high  $Re$, especially in the last two decades \cite{eckhardt2004JFM,kerswell2004JFM,kerswell2005Nonlinear,kerswell2018nonlinear,mullin2006PRL,mullin2011ARFM,hof2006nature,hof2008PRL,hof2011science,barkley2016,hof2023ARFM}. While Newtonian pipe flows have long been shown to be linearly stable, their global stability is not guaranteed. In fact, despite their proven linear stability, Newtonian parallel shear flows become unstable to finite-size external perturbations at finite $Re$ \cite{Drazin2004hydrodynamic}. Only about four decades ago it was realized that the discrepancy between the proven linear stability and the experimental observations is explained by the non-Hermitian Orr-Sommerfeld equation, the linearized Navier-Stokes equation for Newtonian parallel shear flows \cite{Drazin2004hydrodynamic}. At finite $Re$, this equation is prone to transient algebraic time growth of non-normal modes, whereas linear eigenmodes decay exponentially, since the flow is linearly stable \cite{Drazin2004hydrodynamic,schmid2007nonmodal,trefethen1993hydrodynamic,kerswell2018nonlinear}. The stochastic nature of the non-normal mode instability, recognized since Reynolds' seminal experiments, has only recently been experimentally demonstrated in viscoelastic parallel shear flows \cite{hof2003prl,mullin2006PRL,barkley2016}. The dependence of $Re_c$ on the external perturbation intensity contradicts the normal mode linear instability onset independent of perturbation intensity \cite{Drazin2004hydrodynamic}. However, the pathway from laminar flow to sustained turbulence via subcritical non-modal instability in Newtonian pipe flow at $Re\gg 1$ by convective and absolute instabilities is significantly different and more complicated \cite{eckhardt2003,kerswell2004JFM,mullin2006PRL,hof2006nature,hof2008PRL,hof2011science,barkley2016,hof2023PRL}, than the pathway from laminar to sustained chaotic flow via  supercritical non-normal-mode elastic absolute instability in viscoelastic channel flows at $Wi\gg 1$, $Re\ll1$ and $El\gg 1$ \cite{jha2020universal,shnapp2022nonmodal, Li2023non-Hermitian,Li2023flowprop,Steinberg_LTP2022}. As suggested in Ref. \cite{Li2023non-Hermitian}, the surprising difference between the flows is due to the presence of only one nonlinear advection term in the Navier-Stokes equation, as opposed to three nonlinear terms in the elastic stress equation, of which only one is an advection term.

\begin{acknowledgments}
We thank Guy Han and Rostyslav Baron for their help with the experimental setup. This work was supported in part by the Israel Science Foundation (ISF, grant \#784/19). 
\end{acknowledgments}

\appendix

\section*{Materials and Methods}

\subsection{Experimental setup and flow discharge measurements} 
The experiments are conducted in a straight channel of 500(\textit{L}) $\times$ 3.5($W$) $\times$ 0.5(\textit{h}) mm$^3$ dimensions, shown in Fig. \ref{fig:1_sch}.
The fluid is driven by N$_2$ gas at pressure up to 10psi. The fluid discharge is weighed instantaneously, $m(t)$, by a PC-interfaced balance (BPS-1000-C2-V2, MRC) to measure the time-averaged fluid discharge rate $Q=\langle\Delta m/\Delta t\rangle$ to get the mean velocity $U=Q/\rho W h$. Then $Wi=\lambda U/ h$ and $Re=\rho Uh/\eta$ vary in the ranges (30, 300) and (0.005, 0.045), respectively. High resolution ($0.1\%$ of full scale) absolute pressure sensors located near the inlet and outlet at $\Delta l/h\approx10$ (HSC series, Honeywell) with range up to 5 psi are used to measure the pressure fluctuations. For Fig. \ref{fig:3_spike} and Fig. \ref{fig:4_u(t)}, the container of the solution is immediately switched to the other $N_2$ supply line with a slightly higher or lower pressure at $t=0$ and just after switching, $Wi$ is calculated the flow discharge when the flow system reaches equilibrium.

\subsection{Polymer solution preparation and characterization} 
As the working fluid, a dilute polymer solution of high molecular weight polyacrylamide (Polysciences, $M_w=18 MDa$) at a concentration of $c=80$ ppm ($c/c*\approx0.4$ with the overlapping polymer concentration of $c*\approx200$ ppm \cite{solutionpreparation}) is prepared using a water-sucrose solvent with a sugar weight fraction of $64\%$. The properties of the solution are as follows: the solution density ($\rho$) is 1320 kg/$m^3$, the solvent viscosity ($\eta_s$) is 0.13 Pa$\cdot$s, and the total solution viscosity ($\eta$) is 0.17 Pa$\cdot$s. The ratio of solvent viscosity to total solution viscosity is $\eta_s/(\eta_s+\eta_p)=0.765$, where $\eta_p$ is the polymer contribution to the solution viscosity. The longest polymer relaxation time ($\lambda$) is 13 seconds, obtained by the stress relaxation method \cite{solutionpreparation}. The result is $El=Wi/Re= \lambda\rho/\eta d^2=6.8\times 10^3$.

\subsection{Imaging system and PIV measurements} 
We perform velocity field measurements at various distances $l/h$ downstream of the inlet using the Particle Image Velocimetry (PIV) method. The PIV setup consists of 3.2$\mu$m latex fluorescent particle tracers of $\sim$1\% w/w concentration (Thermo Scientific) illuminated by a laser sheet of $\approx 50 \mu$m thickness over the central channel plane, i.e. the $x-z$ plane. A high speed camera (Mini WX100 FASTCAM, Photron) has a high spatial resolution and images of tracer pairs are acquired at 200 to 2000 fps. The OpenPIV software \cite{OpenPIV} is used to analyze $u(x,z,t)$ and $w(x,z,t)$ in the 2D $x$-$z$ plane to record data for $\sim\mathcal{O}(15)$ minutes, or $\sim\mathcal{O}(50\lambda)$, for each $Wi$ to obtain sufficient statistics. For the velocity fluctuations in Figs. \ref{fig:3_spike}, \ref{fig:7_uspec}, and \ref{fig:4_u(t)}, we use spatial averaging over narrow windows with a resolution of 256 ($x$) $\times$96 ($z$) pxl$^2$ with a 4x objective, which serves as a single point velocity measurement at the channel center. The window size for PIV is $32\times 32$ pxl$^2$ with 50\% overlap and 200\% search window size. For the velocity profile visualization in Fig. \ref{fig:10_streak1}, we keep the same scale but increase the spatial resolution to 1280$\times$1280 pxl$^2$.

\nocite{*}
\bibliography{Research_ref}

\begin{thebibliography}{59}%
\makeatletter
\providecommand \@ifxundefined [1]{%
 \@ifx{#1\undefined}
}%
\providecommand \@ifnum [1]{%
 \ifnum #1\expandafter \@firstoftwo
 \else \expandafter \@secondoftwo
 \fi
}%
\providecommand \@ifx [1]{%
 \ifx #1\expandafter \@firstoftwo
 \else \expandafter \@secondoftwo
 \fi
}%
\providecommand \natexlab [1]{#1}%
\providecommand \enquote  [1]{``#1''}%
\providecommand \bibnamefont  [1]{#1}%
\providecommand \bibfnamefont [1]{#1}%
\providecommand \citenamefont [1]{#1}%
\providecommand \href@noop [0]{\@secondoftwo}%
\providecommand \href [0]{\begingroup \@sanitize@url \@href}%
\providecommand \@href[1]{\@@startlink{#1}\@@href}%
\providecommand \@@href[1]{\endgroup#1\@@endlink}%
\providecommand \@sanitize@url [0]{\catcode `\\12\catcode `\$12\catcode `\&12\catcode `\#12\catcode `\^12\catcode `\_12\catcode `\%12\relax}%
\providecommand \@@startlink[1]{}%
\providecommand \@@endlink[0]{}%
\providecommand \url  [0]{\begingroup\@sanitize@url \@url }%
\providecommand \@url [1]{\endgroup\@href {#1}{\urlprefix }}%
\providecommand \urlprefix  [0]{URL }%
\providecommand \Eprint [0]{\href }%
\providecommand \doibase [0]{https://doi.org/}%
\providecommand \selectlanguage [0]{\@gobble}%
\providecommand \bibinfo  [0]{\@secondoftwo}%
\providecommand \bibfield  [0]{\@secondoftwo}%
\providecommand \translation [1]{[#1]}%
\providecommand \BibitemOpen [0]{}%
\providecommand \bibitemStop [0]{}%
\providecommand \bibitemNoStop [0]{.\EOS\space}%
\providecommand \EOS [0]{\spacefactor3000\relax}%
\providecommand \BibitemShut  [1]{\csname bibitem#1\endcsname}%
\let\auto@bib@innerbib\@empty
\bibitem [{\citenamefont {Drazin}\ and\ \citenamefont {Reid}(2004)}]{Drazin2004hydrodynamic}%
  \BibitemOpen
  \bibfield  {author} {\bibinfo {author} {\bibfnamefont {P.~G.}\ \bibnamefont {Drazin}}\ and\ \bibinfo {author} {\bibfnamefont {W.~H.}\ \bibnamefont {Reid}},\ }\href@noop {} {\emph {\bibinfo {title} {Hydrodynamic stability}}}\ (\bibinfo  {publisher} {Cambridge university press},\ \bibinfo {year} {2004})\BibitemShut {NoStop}%
\bibitem [{\citenamefont {Larson}\ \emph {et~al.}(1990)\citenamefont {Larson}, \citenamefont {Shaqfeh},\ and\ \citenamefont {Muller}}]{Larson_JFM1990}%
  \BibitemOpen
  \bibfield  {author} {\bibinfo {author} {\bibfnamefont {R.~G.}\ \bibnamefont {Larson}}, \bibinfo {author} {\bibfnamefont {E.~S.}\ \bibnamefont {Shaqfeh}},\ and\ \bibinfo {author} {\bibfnamefont {S.~J.}\ \bibnamefont {Muller}},\ }\bibfield  {title} {\bibinfo {title} {A purely elastic instability in taylor--couette flow},\ }\href@noop {} {\bibfield  {journal} {\bibinfo  {journal} {J. Fluid Mech.}\ }\textbf {\bibinfo {volume} {218}},\ \bibinfo {pages} {573} (\bibinfo {year} {1990})}\BibitemShut {NoStop}%
\bibitem [{\citenamefont {Shaqfeh}(1996)}]{shaqfeh1996purely}%
  \BibitemOpen
  \bibfield  {author} {\bibinfo {author} {\bibfnamefont {E.~S.}\ \bibnamefont {Shaqfeh}},\ }\bibfield  {title} {\bibinfo {title} {Purely elastic instabilities in viscometric flows},\ }\href@noop {} {\bibfield  {journal} {\bibinfo  {journal} {Annu. Rev. Fluid Mech.}\ }\textbf {\bibinfo {volume} {28}},\ \bibinfo {pages} {129} (\bibinfo {year} {1996})}\BibitemShut {NoStop}%
\bibitem [{\citenamefont {Steinberg}(2021)}]{Steinberg_ARFM2021}%
  \BibitemOpen
  \bibfield  {author} {\bibinfo {author} {\bibfnamefont {V.}~\bibnamefont {Steinberg}},\ }\bibfield  {title} {\bibinfo {title} {Elastic turbulence: an experimental view on inertialess random flow},\ }\href@noop {} {\bibfield  {journal} {\bibinfo  {journal} {Annu. Rev. Fluid Mech.}\ }\textbf {\bibinfo {volume} {53}},\ \bibinfo {pages} {27} (\bibinfo {year} {2021})}\BibitemShut {NoStop}%
\bibitem [{\citenamefont {Schmid}(2007)}]{schmid2007nonmodal}%
  \BibitemOpen
  \bibfield  {author} {\bibinfo {author} {\bibfnamefont {P.~J.}\ \bibnamefont {Schmid}},\ }\bibfield  {title} {\bibinfo {title} {Nonmodal stability theory},\ }\href@noop {} {\bibfield  {journal} {\bibinfo  {journal} {Annu. Rev. Fluid Mech.}\ }\textbf {\bibinfo {volume} {39}},\ \bibinfo {pages} {129} (\bibinfo {year} {2007})}\BibitemShut {NoStop}%
\bibitem [{\citenamefont {Trefethen}\ \emph {et~al.}(1993)\citenamefont {Trefethen}, \citenamefont {Trefethen}, \citenamefont {Reddy},\ and\ \citenamefont {Driscoll}}]{trefethen1993hydrodynamic}%
  \BibitemOpen
  \bibfield  {author} {\bibinfo {author} {\bibfnamefont {L.~N.}\ \bibnamefont {Trefethen}}, \bibinfo {author} {\bibfnamefont {A.~E.}\ \bibnamefont {Trefethen}}, \bibinfo {author} {\bibfnamefont {S.~C.}\ \bibnamefont {Reddy}},\ and\ \bibinfo {author} {\bibfnamefont {T.~A.}\ \bibnamefont {Driscoll}},\ }\bibfield  {title} {\bibinfo {title} {Hydrodynamic stability without eigenvalues},\ }\href@noop {} {\bibfield  {journal} {\bibinfo  {journal} {Science}\ }\textbf {\bibinfo {volume} {261}},\ \bibinfo {pages} {578} (\bibinfo {year} {1993})}\BibitemShut {NoStop}%
\bibitem [{\citenamefont {Kerswell}(2018)}]{kerswell2018nonlinear}%
  \BibitemOpen
  \bibfield  {author} {\bibinfo {author} {\bibfnamefont {R.}~\bibnamefont {Kerswell}},\ }\bibfield  {title} {\bibinfo {title} {Nonlinear nonmodal stability theory},\ }\href@noop {} {\bibfield  {journal} {\bibinfo  {journal} {Annu. Rev. Fluid Mech.}\ }\textbf {\bibinfo {volume} {50}},\ \bibinfo {pages} {319} (\bibinfo {year} {2018})}\BibitemShut {NoStop}%
\bibitem [{\citenamefont {Cross}(1993)}]{cross1993revmodphys}%
  \BibitemOpen
  \bibfield  {author} {\bibinfo {author} {\bibfnamefont {P.~C.}\ \bibnamefont {Cross}, \bibfnamefont {M.~C. \&~Hohenberg}},\ }\bibfield  {title} {\bibinfo {title} {Pattern formation outside of equilibrium},\ }\href@noop {} {\bibfield  {journal} {\bibinfo  {journal} {Rev. Mod. Phys.}\ }\textbf {\bibinfo {volume} {65}},\ \bibinfo {pages} {851} (\bibinfo {year} {1993})}\BibitemShut {NoStop}%
\bibitem [{\citenamefont {Pakdel}\ and\ \citenamefont {McKinley}(1996)}]{pakdel1996elastic}%
  \BibitemOpen
  \bibfield  {author} {\bibinfo {author} {\bibfnamefont {P.}~\bibnamefont {Pakdel}}\ and\ \bibinfo {author} {\bibfnamefont {G.~H.}\ \bibnamefont {McKinley}},\ }\bibfield  {title} {\bibinfo {title} {Elastic instability and curved streamlines},\ }\href@noop {} {\bibfield  {journal} {\bibinfo  {journal} {Phys. Rev. Lett.}\ }\textbf {\bibinfo {volume} {77}},\ \bibinfo {pages} {2459} (\bibinfo {year} {1996})}\BibitemShut {NoStop}%
\bibitem [{\citenamefont {Gorodtsov}\ and\ \citenamefont {Leonov}(1967)}]{Leonov_JMM1967}%
  \BibitemOpen
  \bibfield  {author} {\bibinfo {author} {\bibfnamefont {V.}~\bibnamefont {Gorodtsov}}\ and\ \bibinfo {author} {\bibfnamefont {A.}~\bibnamefont {Leonov}},\ }\bibfield  {title} {\bibinfo {title} {On a linear instability of a plane parallel couette flow of viscoelastic fluid},\ }\href@noop {} {\bibfield  {journal} {\bibinfo  {journal} {J. Appl. Math. Mech.}\ }\textbf {\bibinfo {volume} {31}},\ \bibinfo {pages} {310} (\bibinfo {year} {1967})}\BibitemShut {NoStop}%
\bibitem [{\citenamefont {Renardy}\ and\ \citenamefont {Renardy}(1986)}]{Renardy_JNFM1986}%
  \BibitemOpen
  \bibfield  {author} {\bibinfo {author} {\bibfnamefont {M.}~\bibnamefont {Renardy}}\ and\ \bibinfo {author} {\bibfnamefont {Y.}~\bibnamefont {Renardy}},\ }\bibfield  {title} {\bibinfo {title} {Linear stability of plane couette flow of an upper convected maxwell fluid},\ }\href@noop {} {\bibfield  {journal} {\bibinfo  {journal} {J. Nonnewton. Fluid Mech.}\ }\textbf {\bibinfo {volume} {22}},\ \bibinfo {pages} {23} (\bibinfo {year} {1986})}\BibitemShut {NoStop}%
\bibitem [{\citenamefont {S{\'a}nches}\ \emph {et~al.}(2022)\citenamefont {S{\'a}nches}, \citenamefont {Jovanovi{\'c}}, \citenamefont {Kumar}, \citenamefont {Morozov}, \citenamefont {Shankar}, \citenamefont {Subramanian},\ and\ \citenamefont {Wilson}}]{morozov2022JNNFM}%
  \BibitemOpen
  \bibfield  {author} {\bibinfo {author} {\bibfnamefont {H.~A.~C.}\ \bibnamefont {S{\'a}nches}}, \bibinfo {author} {\bibfnamefont {M.~R.}\ \bibnamefont {Jovanovi{\'c}}}, \bibinfo {author} {\bibfnamefont {S.}~\bibnamefont {Kumar}}, \bibinfo {author} {\bibfnamefont {A.}~\bibnamefont {Morozov}}, \bibinfo {author} {\bibfnamefont {V.}~\bibnamefont {Shankar}}, \bibinfo {author} {\bibfnamefont {G.}~\bibnamefont {Subramanian}},\ and\ \bibinfo {author} {\bibfnamefont {H.~J.}\ \bibnamefont {Wilson}},\ }\bibfield  {title} {\bibinfo {title} {Understanding viscoelastic flow instabilities: Oldroyd-b and beyond},\ }\href@noop {} {\bibfield  {journal} {\bibinfo  {journal} {J. Nonnewton. Fluid Mech.}\ }\textbf {\bibinfo {volume} {302}},\ \bibinfo {pages} {104742} (\bibinfo {year} {2022})}\BibitemShut {NoStop}%
\bibitem [{\citenamefont {Jovanovi{\'c}}\ and\ \citenamefont {Kumar}(2010)}]{jovanovic2010transient}%
  \BibitemOpen
  \bibfield  {author} {\bibinfo {author} {\bibfnamefont {M.~R.}\ \bibnamefont {Jovanovi{\'c}}}\ and\ \bibinfo {author} {\bibfnamefont {S.}~\bibnamefont {Kumar}},\ }\bibfield  {title} {\bibinfo {title} {Transient growth without inertia},\ }\href@noop {} {\bibfield  {journal} {\bibinfo  {journal} {Phys. Fluid}\ }\textbf {\bibinfo {volume} {22}},\ \bibinfo {pages} {023101} (\bibinfo {year} {2010})}\BibitemShut {NoStop}%
\bibitem [{\citenamefont {Jovanovi{\'c}}\ and\ \citenamefont {Kumar}(2011)}]{jovanovic2011nonmodal}%
  \BibitemOpen
  \bibfield  {author} {\bibinfo {author} {\bibfnamefont {M.~R.}\ \bibnamefont {Jovanovi{\'c}}}\ and\ \bibinfo {author} {\bibfnamefont {S.}~\bibnamefont {Kumar}},\ }\bibfield  {title} {\bibinfo {title} {Nonmodal amplification of stochastic disturbances in strongly elastic channel flows},\ }\href@noop {} {\bibfield  {journal} {\bibinfo  {journal} {J. Nonnewton. Fluid Mech.}\ }\textbf {\bibinfo {volume} {166}},\ \bibinfo {pages} {755} (\bibinfo {year} {2011})}\BibitemShut {NoStop}%
\bibitem [{\citenamefont {Page}\ and\ \citenamefont {Zaki}(2014)}]{page2014streak}%
  \BibitemOpen
  \bibfield  {author} {\bibinfo {author} {\bibfnamefont {J.}~\bibnamefont {Page}}\ and\ \bibinfo {author} {\bibfnamefont {T.~A.}\ \bibnamefont {Zaki}},\ }\bibfield  {title} {\bibinfo {title} {Streak evolution in viscoelastic couette flow},\ }\href@noop {} {\bibfield  {journal} {\bibinfo  {journal} {J. Fluid Mech.}\ }\textbf {\bibinfo {volume} {742}},\ \bibinfo {pages} {520} (\bibinfo {year} {2014})}\BibitemShut {NoStop}%
\bibitem [{\citenamefont {Hariharan}\ \emph {et~al.}(2021)\citenamefont {Hariharan}, \citenamefont {Jovanovi{\'c}},\ and\ \citenamefont {Kumar}}]{hariharan2021localized}%
  \BibitemOpen
  \bibfield  {author} {\bibinfo {author} {\bibfnamefont {G.}~\bibnamefont {Hariharan}}, \bibinfo {author} {\bibfnamefont {M.~R.}\ \bibnamefont {Jovanovi{\'c}}},\ and\ \bibinfo {author} {\bibfnamefont {S.}~\bibnamefont {Kumar}},\ }\bibfield  {title} {\bibinfo {title} {Localized stress amplification in inertialess channel flows of viscoelastic fluids},\ }\href@noop {} {\bibfield  {journal} {\bibinfo  {journal} {J. Nonnewton. Fluid Mech.}\ }\textbf {\bibinfo {volume} {291}},\ \bibinfo {pages} {104514} (\bibinfo {year} {2021})}\BibitemShut {NoStop}%
\bibitem [{\citenamefont {Lieu}\ \emph {et~al.}(2013)\citenamefont {Lieu}, \citenamefont {Jovanovi{\'c}},\ and\ \citenamefont {Kumar}}]{lieu2013worst}%
  \BibitemOpen
  \bibfield  {author} {\bibinfo {author} {\bibfnamefont {B.~K.}\ \bibnamefont {Lieu}}, \bibinfo {author} {\bibfnamefont {M.~R.}\ \bibnamefont {Jovanovi{\'c}}},\ and\ \bibinfo {author} {\bibfnamefont {S.}~\bibnamefont {Kumar}},\ }\bibfield  {title} {\bibinfo {title} {Worst-case amplification of disturbances in inertialess couette flow of viscoelastic fluids},\ }\href@noop {} {\bibfield  {journal} {\bibinfo  {journal} {J. Fluid Mech.}\ }\textbf {\bibinfo {volume} {723}},\ \bibinfo {pages} {232} (\bibinfo {year} {2013})}\BibitemShut {NoStop}%
\bibitem [{\citenamefont {Yesilata}(2002)}]{yesilata2002fluiddyn}%
  \BibitemOpen
  \bibfield  {author} {\bibinfo {author} {\bibfnamefont {B.}~\bibnamefont {Yesilata}},\ }\bibfield  {title} {\bibinfo {title} {Nonlinear dynamics of a highly viscous and elastic fluid in pipe flow},\ }\href@noop {} {\bibfield  {journal} {\bibinfo  {journal} {Fluid Dyn. Res.}\ }\textbf {\bibinfo {volume} {31}},\ \bibinfo {pages} {41} (\bibinfo {year} {2002})}\BibitemShut {NoStop}%
\bibitem [{\citenamefont {Yesilata}(2009)}]{yesilata2009polymereng}%
  \BibitemOpen
  \bibfield  {author} {\bibinfo {author} {\bibfnamefont {B.}~\bibnamefont {Yesilata}},\ }\bibfield  {title} {\bibinfo {title} {Temporal nature of polymeric flows near circular pipe-exit},\ }\href@noop {} {\bibfield  {journal} {\bibinfo  {journal} {Polym-Plast. Technol.}\ }\textbf {\bibinfo {volume} {48}},\ \bibinfo {pages} {723} (\bibinfo {year} {2009})}\BibitemShut {NoStop}%
\bibitem [{\citenamefont {Bonn}\ \emph {et~al.}(2011)\citenamefont {Bonn}, \citenamefont {Ingremeau}, \citenamefont {Amarouchene},\ and\ \citenamefont {Kellay}}]{bonn2011large}%
  \BibitemOpen
  \bibfield  {author} {\bibinfo {author} {\bibfnamefont {D.}~\bibnamefont {Bonn}}, \bibinfo {author} {\bibfnamefont {F.}~\bibnamefont {Ingremeau}}, \bibinfo {author} {\bibfnamefont {Y.}~\bibnamefont {Amarouchene}},\ and\ \bibinfo {author} {\bibfnamefont {H.}~\bibnamefont {Kellay}},\ }\bibfield  {title} {\bibinfo {title} {Large velocity fluctuations in small-reynolds-number pipe flow of polymer solutions},\ }\href@noop {} {\bibfield  {journal} {\bibinfo  {journal} {Phys. Rev. E}\ }\textbf {\bibinfo {volume} {84}},\ \bibinfo {pages} {045301} (\bibinfo {year} {2011})}\BibitemShut {NoStop}%
\bibitem [{\citenamefont {Pan}\ \emph {et~al.}(2013)\citenamefont {Pan}, \citenamefont {Morozov}, \citenamefont {Wagner},\ and\ \citenamefont {Arratia}}]{pan2013nonlinear}%
  \BibitemOpen
  \bibfield  {author} {\bibinfo {author} {\bibfnamefont {L.}~\bibnamefont {Pan}}, \bibinfo {author} {\bibfnamefont {A.}~\bibnamefont {Morozov}}, \bibinfo {author} {\bibfnamefont {C.}~\bibnamefont {Wagner}},\ and\ \bibinfo {author} {\bibfnamefont {P.}~\bibnamefont {Arratia}},\ }\bibfield  {title} {\bibinfo {title} {Nonlinear elastic instability in channel flows at low reynolds numbers},\ }\href@noop {} {\bibfield  {journal} {\bibinfo  {journal} {Phys. Rev. Lett.}\ }\textbf {\bibinfo {volume} {110}},\ \bibinfo {pages} {174502} (\bibinfo {year} {2013})}\BibitemShut {NoStop}%
\bibitem [{\citenamefont {Qin}\ and\ \citenamefont {Arratia}(2017)}]{qin2017characterizing}%
  \BibitemOpen
  \bibfield  {author} {\bibinfo {author} {\bibfnamefont {B.}~\bibnamefont {Qin}}\ and\ \bibinfo {author} {\bibfnamefont {P.~E.}\ \bibnamefont {Arratia}},\ }\bibfield  {title} {\bibinfo {title} {Characterizing elastic turbulence in channel flows at low reynolds number},\ }\href@noop {} {\bibfield  {journal} {\bibinfo  {journal} {Phys. Rev. Fluids}\ }\textbf {\bibinfo {volume} {2}},\ \bibinfo {pages} {083302} (\bibinfo {year} {2017})}\BibitemShut {NoStop}%
\bibitem [{\citenamefont {Qin}\ \emph {et~al.}(2019)\citenamefont {Qin}, \citenamefont {Salipante}, \citenamefont {Hudson},\ and\ \citenamefont {Arratia}}]{qin2019flow}%
  \BibitemOpen
  \bibfield  {author} {\bibinfo {author} {\bibfnamefont {B.}~\bibnamefont {Qin}}, \bibinfo {author} {\bibfnamefont {P.~F.}\ \bibnamefont {Salipante}}, \bibinfo {author} {\bibfnamefont {S.~D.}\ \bibnamefont {Hudson}},\ and\ \bibinfo {author} {\bibfnamefont {P.~E.}\ \bibnamefont {Arratia}},\ }\bibfield  {title} {\bibinfo {title} {Flow resistance and structures in viscoelastic channel flows at low re},\ }\href@noop {} {\bibfield  {journal} {\bibinfo  {journal} {Phys. Rev. Lett.}\ }\textbf {\bibinfo {volume} {123}},\ \bibinfo {pages} {194501} (\bibinfo {year} {2019})}\BibitemShut {NoStop}%
\bibitem [{\citenamefont {Jha}\ and\ \citenamefont {Steinberg}(2020)}]{jha2020universal}%
  \BibitemOpen
  \bibfield  {author} {\bibinfo {author} {\bibfnamefont {N.~K.}\ \bibnamefont {Jha}}\ and\ \bibinfo {author} {\bibfnamefont {V.}~\bibnamefont {Steinberg}},\ }\href@noop {} {\bibinfo {title} {Universal coherent structures of elastic turbulence in straight channel with viscoelastic fluid flow}} (\bibinfo {year} {2020}),\ \bibinfo {note} {arXiv preprint https://arxiv.org/abs/2009.12258}\BibitemShut {NoStop}%
\bibitem [{\citenamefont {Jha}\ and\ \citenamefont {Steinberg}(2021)}]{jha2021elastically}%
  \BibitemOpen
  \bibfield  {author} {\bibinfo {author} {\bibfnamefont {N.~K.}\ \bibnamefont {Jha}}\ and\ \bibinfo {author} {\bibfnamefont {V.}~\bibnamefont {Steinberg}},\ }\bibfield  {title} {\bibinfo {title} {Elastically driven kelvin--helmholtz-like instability in straight channel flow},\ }\href@noop {} {\bibfield  {journal} {\bibinfo  {journal} {Proceedings of the National Academy of Sciences}\ }\textbf {\bibinfo {volume} {118}},\ \bibinfo {pages} {e2105211118} (\bibinfo {year} {2021})}\BibitemShut {NoStop}%
\bibitem [{\citenamefont {Shnapp}\ and\ \citenamefont {Steinberg}(2022)}]{shnapp2022nonmodal}%
  \BibitemOpen
  \bibfield  {author} {\bibinfo {author} {\bibfnamefont {R.}~\bibnamefont {Shnapp}}\ and\ \bibinfo {author} {\bibfnamefont {V.}~\bibnamefont {Steinberg}},\ }\bibfield  {title} {\bibinfo {title} {Nonmodal elastic instability and elastic waves in weakly perturbed channel flow},\ }\href@noop {} {\bibfield  {journal} {\bibinfo  {journal} {Phys. Rev. Fluids}\ }\textbf {\bibinfo {volume} {7}},\ \bibinfo {pages} {063901} (\bibinfo {year} {2022})}\BibitemShut {NoStop}%
\bibitem [{\citenamefont {Steinberg}(2022)}]{Steinberg_LTP2022}%
  \BibitemOpen
  \bibfield  {author} {\bibinfo {author} {\bibfnamefont {V.}~\bibnamefont {Steinberg}},\ }\bibfield  {title} {\bibinfo {title} {New direction and perspectives in elastic instability and turbulence in various viscoelastic flow geometries without inertia},\ }\href@noop {} {\bibfield  {journal} {\bibinfo  {journal} {Low Temp. Phys.}\ }\textbf {\bibinfo {volume} {48}},\ \bibinfo {pages} {492} (\bibinfo {year} {2022})}\BibitemShut {NoStop}%
\bibitem [{\citenamefont {Li}\ and\ \citenamefont {Steinberg}(2023{\natexlab{a}})}]{Li2023flowprop}%
  \BibitemOpen
  \bibfield  {author} {\bibinfo {author} {\bibfnamefont {Y.}~\bibnamefont {Li}}\ and\ \bibinfo {author} {\bibfnamefont {V.}~\bibnamefont {Steinberg}},\ }\bibfield  {title} {\bibinfo {title} {Mechanism of vorticity amplification by elastic waves in a viscoelastic channel flow},\ }\href@noop {} {\bibfield  {journal} {\bibinfo  {journal} {Proc. Nat. Acad. Sci. U. S. A.}\ }\textbf {\bibinfo {volume} {120}},\ \bibinfo {pages} {e2305595120} (\bibinfo {year} {2023}{\natexlab{a}})}\BibitemShut {NoStop}%
\bibitem [{\citenamefont {Li}\ and\ \citenamefont {Steinberg}(2023{\natexlab{b}})}]{Li2023non-Hermitian}%
  \BibitemOpen
  \bibfield  {author} {\bibinfo {author} {\bibfnamefont {Y.}~\bibnamefont {Li}}\ and\ \bibinfo {author} {\bibfnamefont {V.}~\bibnamefont {Steinberg}},\ }\bibfield  {title} {\bibinfo {title} {Elastic instability in a straight channel of viscoelastic flow without prearranged perturbations},\ }\href@noop {} {\bibfield  {journal} {\bibinfo  {journal} {Sci. Rep.}\ }\textbf {\bibinfo {volume} {13}},\ \bibinfo {pages} {1064} (\bibinfo {year} {2023}{\natexlab{b}})}\BibitemShut {NoStop}%
\bibitem [{\citenamefont {Datta}\ \emph {et~al.}(2022)\citenamefont {Datta}, \citenamefont {Ardekani}, \citenamefont {Arratia}, \citenamefont {Beris}, \citenamefont {Bischofberger}, \citenamefont {McKinley}, \citenamefont {Eggers}, \citenamefont {L{\'o}pez-Aguilar}, \citenamefont {Fielding}, \citenamefont {Frishman}, \citenamefont {Graham}, \citenamefont {Guasto}, \citenamefont {Haward}, \citenamefont {Shen}, \citenamefont {Hormozi}, \citenamefont {Morozov}, \citenamefont {Poole}, \citenamefont {Shankar}, \citenamefont {Shaqfeh}, \citenamefont {Stark}, \citenamefont {Steinberg}, \citenamefont {Subramanian},\ and\ \citenamefont {Stone}}]{datta2022PRF}%
  \BibitemOpen
  \bibfield  {author} {\bibinfo {author} {\bibfnamefont {S.~S.}\ \bibnamefont {Datta}}, \bibinfo {author} {\bibfnamefont {A.~M.}\ \bibnamefont {Ardekani}}, \bibinfo {author} {\bibfnamefont {P.~E.}\ \bibnamefont {Arratia}}, \bibinfo {author} {\bibfnamefont {A.~N.}\ \bibnamefont {Beris}}, \bibinfo {author} {\bibfnamefont {I.}~\bibnamefont {Bischofberger}}, \bibinfo {author} {\bibfnamefont {G.~H.}\ \bibnamefont {McKinley}}, \bibinfo {author} {\bibfnamefont {J.~G.}\ \bibnamefont {Eggers}}, \bibinfo {author} {\bibfnamefont {J.~E.}\ \bibnamefont {L{\'o}pez-Aguilar}}, \bibinfo {author} {\bibfnamefont {S.~M.}\ \bibnamefont {Fielding}}, \bibinfo {author} {\bibfnamefont {A.}~\bibnamefont {Frishman}}, \bibinfo {author} {\bibfnamefont {M.~D.}\ \bibnamefont {Graham}}, \bibinfo {author} {\bibfnamefont {J.~S.}\ \bibnamefont {Guasto}}, \bibinfo {author} {\bibfnamefont {S.~J.}\ \bibnamefont {Haward}}, \bibinfo {author} {\bibfnamefont {A.~Q.}\ \bibnamefont {Shen}}, \bibinfo {author} {\bibfnamefont {S.}~\bibnamefont {Hormozi}},
  \bibinfo {author} {\bibfnamefont {A.}~\bibnamefont {Morozov}}, \bibinfo {author} {\bibfnamefont {R.~J.}\ \bibnamefont {Poole}}, \bibinfo {author} {\bibfnamefont {V.}~\bibnamefont {Shankar}}, \bibinfo {author} {\bibfnamefont {E.~S.~G.}\ \bibnamefont {Shaqfeh}}, \bibinfo {author} {\bibfnamefont {H.}~\bibnamefont {Stark}}, \bibinfo {author} {\bibfnamefont {V.}~\bibnamefont {Steinberg}}, \bibinfo {author} {\bibfnamefont {G.}~\bibnamefont {Subramanian}},\ and\ \bibinfo {author} {\bibfnamefont {H.~A.}\ \bibnamefont {Stone}},\ }\bibfield  {title} {\bibinfo {title} {Perspectives on viscoelatic flow instabilities and elastic turbulence},\ }\href@noop {} {\bibfield  {journal} {\bibinfo  {journal} {Phys. Rev. Fluids}\ }\textbf {\bibinfo {volume} {7}},\ \bibinfo {pages} {080701} (\bibinfo {year} {2022})}\BibitemShut {NoStop}%
\bibitem [{\citenamefont {Balkovsky}\ \emph {et~al.}(2001)\citenamefont {Balkovsky}, \citenamefont {Fouxon},\ and\ \citenamefont {Lebedev}}]{Lebedev2001PRE}%
  \BibitemOpen
  \bibfield  {author} {\bibinfo {author} {\bibfnamefont {E.}~\bibnamefont {Balkovsky}}, \bibinfo {author} {\bibfnamefont {A.}~\bibnamefont {Fouxon}},\ and\ \bibinfo {author} {\bibfnamefont {V.}~\bibnamefont {Lebedev}},\ }\bibfield  {title} {\bibinfo {title} {Turbulence of polymer solutions},\ }\href@noop {} {\bibfield  {journal} {\bibinfo  {journal} {Phys. Rev. E}\ }\textbf {\bibinfo {volume} {64}},\ \bibinfo {pages} {056301} (\bibinfo {year} {2001})}\BibitemShut {NoStop}%
\bibitem [{\citenamefont {Varshney}\ and\ \citenamefont {Steinberg}(2019)}]{varshney2019NatComm}%
  \BibitemOpen
  \bibfield  {author} {\bibinfo {author} {\bibfnamefont {A.}~\bibnamefont {Varshney}}\ and\ \bibinfo {author} {\bibfnamefont {V.}~\bibnamefont {Steinberg}},\ }\bibfield  {title} {\bibinfo {title} {Elastic alfven waves in elastic turbulence},\ }\href@noop {} {\bibfield  {journal} {\bibinfo  {journal} {Nat. Commun.}\ }\textbf {\bibinfo {volume} {10}},\ \bibinfo {pages} {652} (\bibinfo {year} {2019})}\BibitemShut {NoStop}%
\bibitem [{\citenamefont {Varshney}\ and\ \citenamefont {Steinberg}(2018)}]{varshney2018PRF}%
  \BibitemOpen
  \bibfield  {author} {\bibinfo {author} {\bibfnamefont {A.}~\bibnamefont {Varshney}}\ and\ \bibinfo {author} {\bibfnamefont {V.}~\bibnamefont {Steinberg}},\ }\bibfield  {title} {\bibinfo {title} {Drag enhancement and drag reduction in viscoelastic flow},\ }\href@noop {} {\bibfield  {journal} {\bibinfo  {journal} {Phys. Rev. Fluids}\ }\textbf {\bibinfo {volume} {3}},\ \bibinfo {pages} {103302} (\bibinfo {year} {2018})}\BibitemShut {NoStop}%
\bibitem [{\citenamefont {Kumar}\ \emph {et~al.}(2022)\citenamefont {Kumar}, \citenamefont {Varshney}, \citenamefont {Li},\ and\ \citenamefont {Steinberg}}]{kumar2022PRF}%
  \BibitemOpen
  \bibfield  {author} {\bibinfo {author} {\bibfnamefont {M.~V.}\ \bibnamefont {Kumar}}, \bibinfo {author} {\bibfnamefont {A.}~\bibnamefont {Varshney}}, \bibinfo {author} {\bibfnamefont {D.}~\bibnamefont {Li}},\ and\ \bibinfo {author} {\bibfnamefont {V.}~\bibnamefont {Steinberg}},\ }\bibfield  {title} {\bibinfo {title} {Relaminarization of elastic turbulence},\ }\href@noop {} {\bibfield  {journal} {\bibinfo  {journal} {Phys. Rev. Fluids}\ }\textbf {\bibinfo {volume} {7}},\ \bibinfo {pages} {L081301} (\bibinfo {year} {2022})}\BibitemShut {NoStop}%
\bibitem [{\citenamefont {Anishchenko}\ \emph {et~al.}(1999)\citenamefont {Anishchenko}, \citenamefont {Neiman}, \citenamefont {Moss},\ and\ \citenamefont {Shimansky-Geier}}]{anishch1999phys_uspekhi}%
  \BibitemOpen
  \bibfield  {author} {\bibinfo {author} {\bibfnamefont {V.~S.}\ \bibnamefont {Anishchenko}}, \bibinfo {author} {\bibfnamefont {A.~B.}\ \bibnamefont {Neiman}}, \bibinfo {author} {\bibfnamefont {F.}~\bibnamefont {Moss}},\ and\ \bibinfo {author} {\bibfnamefont {L.}~\bibnamefont {Shimansky-Geier}},\ }\bibfield  {title} {\bibinfo {title} {Stochastic resonance: noise-enhanced order},\ }\href@noop {} {\bibfield  {journal} {\bibinfo  {journal} {Physics-Uspekhi}\ }\textbf {\bibinfo {volume} {42}},\ \bibinfo {pages} {7} (\bibinfo {year} {1999})}\BibitemShut {NoStop}%
\bibitem [{\citenamefont {Anishchenko}\ \emph {et~al.}(1993)\citenamefont {Anishchenko}, \citenamefont {Neiman},\ and\ \citenamefont {Safanova}}]{anishch1993jstatphys}%
  \BibitemOpen
  \bibfield  {author} {\bibinfo {author} {\bibfnamefont {V.~S.}\ \bibnamefont {Anishchenko}}, \bibinfo {author} {\bibfnamefont {A.~B.}\ \bibnamefont {Neiman}},\ and\ \bibinfo {author} {\bibfnamefont {M.~A.}\ \bibnamefont {Safanova}},\ }\bibfield  {title} {\bibinfo {title} {Stochastic resonance in chaotic systems},\ }\href@noop {} {\bibfield  {journal} {\bibinfo  {journal} {J. Stat. Phys.}\ }\textbf {\bibinfo {volume} {70}},\ \bibinfo {pages} {183} (\bibinfo {year} {1993})}\BibitemShut {NoStop}%
\bibitem [{\citenamefont {Benzi}\ \emph {et~al.}(1981)\citenamefont {Benzi}, \citenamefont {Sutera},\ and\ \citenamefont {Vulpiani}}]{benzi1981JPhys}%
  \BibitemOpen
  \bibfield  {author} {\bibinfo {author} {\bibfnamefont {R.}~\bibnamefont {Benzi}}, \bibinfo {author} {\bibfnamefont {A.}~\bibnamefont {Sutera}},\ and\ \bibinfo {author} {\bibfnamefont {A.}~\bibnamefont {Vulpiani}},\ }\bibfield  {title} {\bibinfo {title} {The mechanism of stochastic resonance},\ }\href@noop {} {\bibfield  {journal} {\bibinfo  {journal} {J. Phys. A: Math. Gen.}\ }\textbf {\bibinfo {volume} {14A}},\ \bibinfo {pages} {L451} (\bibinfo {year} {1981})}\BibitemShut {NoStop}%
\bibitem [{\citenamefont {Schekochihin}(2022)}]{schek2022jplasmaphys}%
  \BibitemOpen
  \bibfield  {author} {\bibinfo {author} {\bibfnamefont {A.~A.}\ \bibnamefont {Schekochihin}},\ }\bibfield  {title} {\bibinfo {title} {Mhd turbulence: a biased review},\ }\href@noop {} {\bibfield  {journal} {\bibinfo  {journal} {J. Plasma Phys.}\ }\textbf {\bibinfo {volume} {88}},\ \bibinfo {pages} {155880501} (\bibinfo {year} {2022})}\BibitemShut {NoStop}%
\bibitem [{\citenamefont {Li}\ \emph {et~al.}(2019)\citenamefont {Li}, \citenamefont {Archer},\ and\ \citenamefont {Koch}}]{koch2019PRL}%
  \BibitemOpen
  \bibfield  {author} {\bibinfo {author} {\bibfnamefont {G.}~\bibnamefont {Li}}, \bibinfo {author} {\bibfnamefont {L.~A.}\ \bibnamefont {Archer}},\ and\ \bibinfo {author} {\bibfnamefont {D.~L.}\ \bibnamefont {Koch}},\ }\bibfield  {title} {\bibinfo {title} {Electroconvection in a viscoelastic electrolyte},\ }\href@noop {} {\bibfield  {journal} {\bibinfo  {journal} {Phys. Rev. Lett.}\ }\textbf {\bibinfo {volume} {122}},\ \bibinfo {pages} {124501} (\bibinfo {year} {2019})}\BibitemShut {NoStop}%
\bibitem [{\citenamefont {Das}\ \emph {et~al.}(2022)\citenamefont {Das}, \citenamefont {Natale},\ and\ \citenamefont {Benneker}}]{das2022jnnfm}%
  \BibitemOpen
  \bibfield  {author} {\bibinfo {author} {\bibfnamefont {S.~K.}\ \bibnamefont {Das}}, \bibinfo {author} {\bibfnamefont {G.}~\bibnamefont {Natale}},\ and\ \bibinfo {author} {\bibfnamefont {A.~M.}\ \bibnamefont {Benneker}},\ }\bibfield  {title} {\bibinfo {title} {Viscoelastic behavior of dilute polyelectrolyte solutions in complex geometries},\ }\href@noop {} {\bibfield  {journal} {\bibinfo  {journal} {J. Nonnewton. Fluid Mech.}\ }\textbf {\bibinfo {volume} {309}},\ \bibinfo {pages} {104920} (\bibinfo {year} {2022})}\BibitemShut {NoStop}%
\bibitem [{\citenamefont {Emmanuel}\ \emph {et~al.}(2017)\citenamefont {Emmanuel}, \citenamefont {Musacchio},\ and\ \citenamefont {Vincenzi}}]{dario2017PRE}%
  \BibitemOpen
  \bibfield  {author} {\bibinfo {author} {\bibfnamefont {L.}~\bibnamefont {Emmanuel}}, \bibinfo {author} {\bibfnamefont {S.}~\bibnamefont {Musacchio}},\ and\ \bibinfo {author} {\bibfnamefont {D.}~\bibnamefont {Vincenzi}},\ }\bibfield  {title} {\bibinfo {title} {Emergence of chaos in a viscous solution of rods},\ }\href@noop {} {\bibfield  {journal} {\bibinfo  {journal} {Phys. Rev. E}\ }\textbf {\bibinfo {volume} {96}},\ \bibinfo {pages} {053108} (\bibinfo {year} {2017})}\BibitemShut {NoStop}%
\bibitem [{\citenamefont {Brouzet}\ \emph {et~al.}(2014)\citenamefont {Brouzet}, \citenamefont {Verhille},\ and\ \citenamefont {Le~Gal}}]{legal2014PRL}%
  \BibitemOpen
  \bibfield  {author} {\bibinfo {author} {\bibfnamefont {C.}~\bibnamefont {Brouzet}}, \bibinfo {author} {\bibfnamefont {G.}~\bibnamefont {Verhille}},\ and\ \bibinfo {author} {\bibfnamefont {P.}~\bibnamefont {Le~Gal}},\ }\bibfield  {title} {\bibinfo {title} {Flexible fiber in a turbulent flow: A macroscopic polymer},\ }\href@noop {} {\bibfield  {journal} {\bibinfo  {journal} {Phys. Rev. Lett.}\ }\textbf {\bibinfo {volume} {112}},\ \bibinfo {pages} {074501} (\bibinfo {year} {2014})}\BibitemShut {NoStop}%
\bibitem [{\citenamefont {Doostmohammadi}\ \emph {et~al.}(2018)\citenamefont {Doostmohammadi}, \citenamefont {Ign{\'e}s-Mullol}, \citenamefont {Yeomans},\ and\ \citenamefont {Sagu{\'e}s}}]{julia2018natcomm}%
  \BibitemOpen
  \bibfield  {author} {\bibinfo {author} {\bibfnamefont {A.}~\bibnamefont {Doostmohammadi}}, \bibinfo {author} {\bibfnamefont {J.}~\bibnamefont {Ign{\'e}s-Mullol}}, \bibinfo {author} {\bibfnamefont {J.~M.}\ \bibnamefont {Yeomans}},\ and\ \bibinfo {author} {\bibfnamefont {F.}~\bibnamefont {Sagu{\'e}s}},\ }\bibfield  {title} {\bibinfo {title} {Active nematics},\ }\href@noop {} {\bibfield  {journal} {\bibinfo  {journal} {Nat. Commun.}\ }\textbf {\bibinfo {volume} {9}},\ \bibinfo {pages} {3246} (\bibinfo {year} {2018})}\BibitemShut {NoStop}%
\bibitem [{\citenamefont {Alert}\ \emph {et~al.}(2022)\citenamefont {Alert}, \citenamefont {Casademunt},\ and\ \citenamefont {Joanny}}]{joanny2022arcmp}%
  \BibitemOpen
  \bibfield  {author} {\bibinfo {author} {\bibfnamefont {R.}~\bibnamefont {Alert}}, \bibinfo {author} {\bibfnamefont {J.}~\bibnamefont {Casademunt}},\ and\ \bibinfo {author} {\bibfnamefont {J.-F.}\ \bibnamefont {Joanny}},\ }\bibfield  {title} {\bibinfo {title} {Active turbulence},\ }\href@noop {} {\bibfield  {journal} {\bibinfo  {journal} {Annual Review of Condensed Matter Physics}\ }\textbf {\bibinfo {volume} {13}},\ \bibinfo {pages} {143} (\bibinfo {year} {2022})}\BibitemShut {NoStop}%
\bibitem [{\citenamefont {Faisst}\ and\ \citenamefont {Eckhardt}(2004)}]{eckhardt2004JFM}%
  \BibitemOpen
  \bibfield  {author} {\bibinfo {author} {\bibfnamefont {H.}~\bibnamefont {Faisst}}\ and\ \bibinfo {author} {\bibfnamefont {B.}~\bibnamefont {Eckhardt}},\ }\bibfield  {title} {\bibinfo {title} {Sensitive dependence on initial conditions in transition to turbulence in pipe flow},\ }\href@noop {} {\bibfield  {journal} {\bibinfo  {journal} {J. Fluid Mech.}\ }\textbf {\bibinfo {volume} {504}},\ \bibinfo {pages} {343} (\bibinfo {year} {2004})}\BibitemShut {NoStop}%
\bibitem [{\citenamefont {Wedin}\ and\ \citenamefont {Kerswell}(2004)}]{kerswell2004JFM}%
  \BibitemOpen
  \bibfield  {author} {\bibinfo {author} {\bibfnamefont {H.}~\bibnamefont {Wedin}}\ and\ \bibinfo {author} {\bibfnamefont {R.~R.}\ \bibnamefont {Kerswell}},\ }\bibfield  {title} {\bibinfo {title} {Exact coherent structures in pipe flow: travelling wave solutions},\ }\href@noop {} {\bibfield  {journal} {\bibinfo  {journal} {J. Fluid Mech.}\ }\textbf {\bibinfo {volume} {508}},\ \bibinfo {pages} {333} (\bibinfo {year} {2004})}\BibitemShut {NoStop}%
\bibitem [{\citenamefont {Kerswell}(2005)}]{kerswell2005Nonlinear}%
  \BibitemOpen
  \bibfield  {author} {\bibinfo {author} {\bibfnamefont {R.}~\bibnamefont {Kerswell}},\ }\bibfield  {title} {\bibinfo {title} {Recent progress in understanding the transition to turbulence in a pipe},\ }\href@noop {} {\bibfield  {journal} {\bibinfo  {journal} {Nonlinearity}\ }\textbf {\bibinfo {volume} {18}},\ \bibinfo {pages} {R17} (\bibinfo {year} {2005})}\BibitemShut {NoStop}%
\bibitem [{\citenamefont {Peixinho}\ and\ \citenamefont {Mullin}(2006)}]{mullin2006PRL}%
  \BibitemOpen
  \bibfield  {author} {\bibinfo {author} {\bibfnamefont {J.}~\bibnamefont {Peixinho}}\ and\ \bibinfo {author} {\bibfnamefont {T.}~\bibnamefont {Mullin}},\ }\bibfield  {title} {\bibinfo {title} {Decay of turbulence in pipe flow},\ }\href@noop {} {\bibfield  {journal} {\bibinfo  {journal} {Phys. Rev. Lett.}\ }\textbf {\bibinfo {volume} {96}},\ \bibinfo {pages} {094501} (\bibinfo {year} {2006})}\BibitemShut {NoStop}%
\bibitem [{\citenamefont {Mullin}(2011)}]{mullin2011ARFM}%
  \BibitemOpen
  \bibfield  {author} {\bibinfo {author} {\bibfnamefont {T.}~\bibnamefont {Mullin}},\ }\bibfield  {title} {\bibinfo {title} {Experimental studies of transition to turbulence in a pipe},\ }\href@noop {} {\bibfield  {journal} {\bibinfo  {journal} {Annu. Rev. Fluid Mech.}\ }\textbf {\bibinfo {volume} {43}},\ \bibinfo {pages} {1} (\bibinfo {year} {2011})}\BibitemShut {NoStop}%
\bibitem [{\citenamefont {Hof}\ \emph {et~al.}(2006)\citenamefont {Hof}, \citenamefont {Westerweel}, \citenamefont {Schneider},\ and\ \citenamefont {Eckhardt}}]{hof2006nature}%
  \BibitemOpen
  \bibfield  {author} {\bibinfo {author} {\bibfnamefont {B.}~\bibnamefont {Hof}}, \bibinfo {author} {\bibfnamefont {J.}~\bibnamefont {Westerweel}}, \bibinfo {author} {\bibfnamefont {T.~M.}\ \bibnamefont {Schneider}},\ and\ \bibinfo {author} {\bibfnamefont {B.}~\bibnamefont {Eckhardt}},\ }\bibfield  {title} {\bibinfo {title} {Finite lifetime of turbulence in shear flows},\ }\href@noop {} {\bibfield  {journal} {\bibinfo  {journal} {Nature (London)}\ }\textbf {\bibinfo {volume} {443}},\ \bibinfo {pages} {59} (\bibinfo {year} {2006})}\BibitemShut {NoStop}%
\bibitem [{\citenamefont {Hof}\ \emph {et~al.}(2008)\citenamefont {Hof}, \citenamefont {Lozar}, \citenamefont {Kuik},\ and\ \citenamefont {Westerweel}}]{hof2008PRL}%
  \BibitemOpen
  \bibfield  {author} {\bibinfo {author} {\bibfnamefont {B.}~\bibnamefont {Hof}}, \bibinfo {author} {\bibfnamefont {A.~D.}\ \bibnamefont {Lozar}}, \bibinfo {author} {\bibfnamefont {D.~J.}\ \bibnamefont {Kuik}},\ and\ \bibinfo {author} {\bibfnamefont {J.}~\bibnamefont {Westerweel}},\ }\bibfield  {title} {\bibinfo {title} {Repeller or attractor? selecting the dynamical model for the onset of turbulence in pipe flow},\ }\href@noop {} {\bibfield  {journal} {\bibinfo  {journal} {Phys. Rev. Lett.}\ }\textbf {\bibinfo {volume} {101}},\ \bibinfo {pages} {214501} (\bibinfo {year} {2008})}\BibitemShut {NoStop}%
\bibitem [{\citenamefont {Avila}\ \emph {et~al.}(2011)\citenamefont {Avila}, \citenamefont {Moxey}, \citenamefont {De~Lozar}, \citenamefont {Avila}, \citenamefont {Barkley},\ and\ \citenamefont {Hof}}]{hof2011science}%
  \BibitemOpen
  \bibfield  {author} {\bibinfo {author} {\bibfnamefont {K.}~\bibnamefont {Avila}}, \bibinfo {author} {\bibfnamefont {D.}~\bibnamefont {Moxey}}, \bibinfo {author} {\bibfnamefont {A.}~\bibnamefont {De~Lozar}}, \bibinfo {author} {\bibfnamefont {M.}~\bibnamefont {Avila}}, \bibinfo {author} {\bibfnamefont {D.}~\bibnamefont {Barkley}},\ and\ \bibinfo {author} {\bibfnamefont {B.}~\bibnamefont {Hof}},\ }\bibfield  {title} {\bibinfo {title} {The onset of turbulence in pipe flow},\ }\href@noop {} {\bibfield  {journal} {\bibinfo  {journal} {Science}\ }\textbf {\bibinfo {volume} {333}},\ \bibinfo {pages} {192} (\bibinfo {year} {2011})}\BibitemShut {NoStop}%
\bibitem [{\citenamefont {Barkley}(2016)}]{barkley2016}%
  \BibitemOpen
  \bibfield  {author} {\bibinfo {author} {\bibfnamefont {D.}~\bibnamefont {Barkley}},\ }\bibfield  {title} {\bibinfo {title} {Theoretical perspective on the route to turbulence in a pipe},\ }\href@noop {} {\bibfield  {journal} {\bibinfo  {journal} {J. Fluid Mech.}\ }\textbf {\bibinfo {volume} {803}},\ \bibinfo {pages} {P1} (\bibinfo {year} {2016})}\BibitemShut {NoStop}%
\bibitem [{\citenamefont {Avila}\ \emph {et~al.}(2023)\citenamefont {Avila}, \citenamefont {Barkley},\ and\ \citenamefont {Hof}}]{hof2023ARFM}%
  \BibitemOpen
  \bibfield  {author} {\bibinfo {author} {\bibfnamefont {M.}~\bibnamefont {Avila}}, \bibinfo {author} {\bibfnamefont {D.}~\bibnamefont {Barkley}},\ and\ \bibinfo {author} {\bibfnamefont {B.}~\bibnamefont {Hof}},\ }\bibfield  {title} {\bibinfo {title} {Transition to turbulence in pipe flow},\ }\href@noop {} {\bibfield  {journal} {\bibinfo  {journal} {Annu. Rev. Fluid Mech.}\ }\textbf {\bibinfo {volume} {55}},\ \bibinfo {pages} {575} (\bibinfo {year} {2023})}\BibitemShut {NoStop}%
\bibitem [{\citenamefont {Hof}\ \emph {et~al.}(2003)\citenamefont {Hof}, \citenamefont {Juel},\ and\ \citenamefont {Mullin}}]{hof2003prl}%
  \BibitemOpen
  \bibfield  {author} {\bibinfo {author} {\bibfnamefont {B.}~\bibnamefont {Hof}}, \bibinfo {author} {\bibfnamefont {A.}~\bibnamefont {Juel}},\ and\ \bibinfo {author} {\bibfnamefont {T.}~\bibnamefont {Mullin}},\ }\bibfield  {title} {\bibinfo {title} {Scaling of the threshold of pipe flow turbulence},\ }\href@noop {} {\bibfield  {journal} {\bibinfo  {journal} {Phys. Rev. Lett.}\ }\textbf {\bibinfo {volume} {91}},\ \bibinfo {pages} {244502} (\bibinfo {year} {2003})}\BibitemShut {NoStop}%
\bibitem [{\citenamefont {Faisst}\ and\ \citenamefont {Eckhardt}(2003)}]{eckhardt2003}%
  \BibitemOpen
  \bibfield  {author} {\bibinfo {author} {\bibfnamefont {H.}~\bibnamefont {Faisst}}\ and\ \bibinfo {author} {\bibfnamefont {B.}~\bibnamefont {Eckhardt}},\ }\bibfield  {title} {\bibinfo {title} {Traveling waves in pipe flow},\ }\href@noop {} {\bibfield  {journal} {\bibinfo  {journal} {Phys. Rev. Lett.}\ }\textbf {\bibinfo {volume} {91}},\ \bibinfo {pages} {224502} (\bibinfo {year} {2003})}\BibitemShut {NoStop}%
\bibitem [{\citenamefont {Paranjape}\ \emph {et~al.}(2023)\citenamefont {Paranjape}, \citenamefont {Yaln\ifmmode \imath \else~\i \fi{}z}, \citenamefont {Duguet}, \citenamefont {Budanur},\ and\ \citenamefont {Hof}}]{hof2023PRL}%
  \BibitemOpen
  \bibfield  {author} {\bibinfo {author} {\bibfnamefont {C.~S.}\ \bibnamefont {Paranjape}}, \bibinfo {author} {\bibfnamefont {G.}~\bibnamefont {Yaln\ifmmode \imath \else~\i \fi{}z}}, \bibinfo {author} {\bibfnamefont {Y.}~\bibnamefont {Duguet}}, \bibinfo {author} {\bibfnamefont {N.~B.}\ \bibnamefont {Budanur}},\ and\ \bibinfo {author} {\bibfnamefont {B.}~\bibnamefont {Hof}},\ }\bibfield  {title} {\bibinfo {title} {Direct path from turbulence to time-periodic solutions},\ }\href {https://doi.org/10.1103/PhysRevLett.131.034002} {\bibfield  {journal} {\bibinfo  {journal} {Phys. Rev. Lett.}\ }\textbf {\bibinfo {volume} {131}},\ \bibinfo {pages} {034002} (\bibinfo {year} {2023})}\BibitemShut {NoStop}%
\bibitem [{\citenamefont {Liu}\ \emph {et~al.}(2009)\citenamefont {Liu}, \citenamefont {Jun},\ and\ \citenamefont {Steinberg}}]{solutionpreparation}%
  \BibitemOpen
  \bibfield  {author} {\bibinfo {author} {\bibfnamefont {Y.}~\bibnamefont {Liu}}, \bibinfo {author} {\bibfnamefont {Y.}~\bibnamefont {Jun}},\ and\ \bibinfo {author} {\bibfnamefont {V.}~\bibnamefont {Steinberg}},\ }\bibfield  {title} {\bibinfo {title} {Concentration dependence of the longest relaxation times of dilute and semi-dilute polymer solutions},\ }\href@noop {} {\bibfield  {journal} {\bibinfo  {journal} {J. Rheol.}\ }\textbf {\bibinfo {volume} {53}},\ \bibinfo {pages} {1069} (\bibinfo {year} {2009})}\BibitemShut {NoStop}%
\bibitem [{\citenamefont {Liberzon}\ \emph {et~al.}(2021)\citenamefont {Liberzon}, \citenamefont {Käufer}, \citenamefont {Bauer}, \citenamefont {Vennemann},\ and\ \citenamefont {Zimmer}}]{OpenPIV}%
  \BibitemOpen
  \bibfield  {author} {\bibinfo {author} {\bibfnamefont {A.}~\bibnamefont {Liberzon}}, \bibinfo {author} {\bibfnamefont {T.}~\bibnamefont {Käufer}}, \bibinfo {author} {\bibfnamefont {A.}~\bibnamefont {Bauer}}, \bibinfo {author} {\bibfnamefont {P.}~\bibnamefont {Vennemann}},\ and\ \bibinfo {author} {\bibfnamefont {E.}~\bibnamefont {Zimmer}},\ }\href {https://doi.org/10.5281/zenodo.5009150} {\bibinfo {title} {Openpiv/openpiv-python: Openpiv-python v0.23.6}} (\bibinfo {year} {2021})\BibitemShut {NoStop}%
\end{thebibliography}%

\end{document}